\documentclass[11pt]{article}
  \usepackage[preprint]{acl}
  \usepackage{times}
  \usepackage{latexsym}
  \usepackage{url}
  \usepackage[utf8]{inputenc}
  \usepackage[T1]{fontenc}
  \usepackage{graphicx}
  \usepackage{amsmath}
  \usepackage{amsthm}
  \usepackage{booktabs}
  \usepackage{algorithm}
  \usepackage{algpseudocode}
  \usepackage{arydshln}
  \usepackage{float}
  \usepackage{multirow}

  \newcommand{\Description}[1]{}

  \title{Beyond Simplification: DFT-GEN for Fidelity-Preserving Visual Accessibility in Dyslexia-Friendly Educational Texts}

  \author{
  Jiaqian Yu$^{1}$ \quad
  Chen Jason Zhang$^{1}$ \quad
  Haoyang Li$^{1}$ \quad
  Guoqiong Ivanka Huang$^{2}$ \\
  $^{1}$Department of Computing, The Hong Kong Polytechnic University \\
  $^{2}$School of Hotel and Tourism Management, The Hong Kong Polytechnic University \\
  \texttt{\small genshin-impact.yu@connect.polyu.hk} \\
  \texttt{\small \{jason-c.zhang, haoyang-comp.li, ivanka.huang\}@polyu.edu.hk}
  }

\begin{document}
  \maketitle

  \begin{abstract}
  Dense educational texts impose avoidable reading friction on people with dyslexia, yet generic simplification can delete terminology, task constraints, or source evidence that readers still need. Stakeholder interviews with dyslexic adults and specialists reveal a core tension: \emph{reduced burden must not compromise information fidelity}. We present \textbf{DFT-GEN}, a stakeholder-informed text transformation framework for content-heavy educational materials. Its central contribution is not a generic LLM refinement loop, but a dyslexia-specific accessibility layer that combines protected-span preservation with a deterministic \textbf{Dyslexia Accessibility Controller} (DAC) for rendered visual organization. DAC converts stakeholder and expert preferences into reproducible controls for visual-unit length, chunk spacing, source/task separation, highlighting budget, and reviewable risk flags. We therefore separate evaluation into DCFI, a fidelity-safety diagnostic, and B-DVAS-VL, a rendered visual-accessibility diagnostic. On 2,280 bilingual exam-style items, DFT-GEN preserves task-critical information while improving visual accessibility: it wins 93\% (EN) and 64\% (ZH) of B-DVAS-VL pairwise judgments against same-backbone controls, and in a controlled pilot with dyslexic adult readers it preserves answerability while reducing effort.
  \end{abstract}

  \section{Introduction}

  Dyslexia affects reading fluency, decoding, and the speed of information processing~\cite{shaywitz2020overcoming,ref_acm_dyslexia_stats}. For many users, the barrier is not conceptual understanding but the interaction between dense presentation, visual stress, and cognitive load. This makes dyslexia accessibility a grounded NLP problem defined by population needs rather than generic simplification metrics.

  Automatic text simplification (ATS) and LLMs appear promising~\cite{chandrasekar1996motivations,xu2016optimizing,agrawal2023controlling,kew2023bless}, but our stakeholder work shows that naive simplification can be harmful. Specialists emphasized that dyslexic readers often understand sophisticated vocabulary but need more time, clearer structure, and visual support~\cite{relloweb,madjidi2024inclusive}. Over-simplification can remove useful terminology or change meaning---especially risky for educational materials where users still need access to original concepts.

  \begin{figure*}[t]
  \centering
  \includegraphics[width=\textwidth]{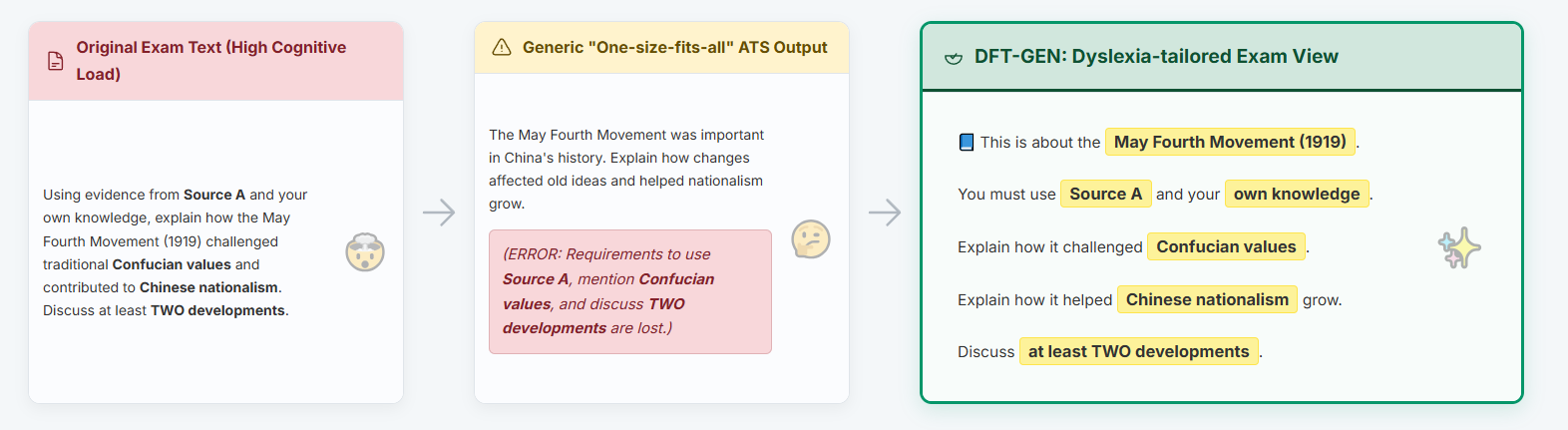}
  \caption{Motivating example. A generic simplifier can make an exam prompt shorter but lose task-critical constraints such as source use, named concepts, and the required number of developments. DFT-GEN instead restructures the prompt while keeping the task anchors visible.}
  \label{fig:hist_motivation}
  \end{figure*}

  We therefore formulate dyslexia-friendly generation as a \emph{constraint-driven accessibility transformation}: reduce avoidable reading friction while preserving the information needed to answer, learn, or act. Three design principles follow directly from the stakeholder evidence, as illustrated by the contrast in Figure~\ref{fig:hist_motivation}. \textbf{P1 (Fidelity first):} preserve key terms, structural markers, numbers, quotations, and instructions unless there is an explicit reason to modify them---because specialists emphasized that a changed text can be more harmful than a difficult one. \textbf{P2 (Visual accessibility as a first-class target):} spacing, segmentation, highlighting, and sentence-level rewriting address different barriers and should be controlled as rendered presentation rather than left to a prompt's style imitation. \textbf{P3 (Reviewability):} deployment should include confidence signals and risk flags so high-impact transformations can be checked by humans.

  These principles also determine what to \emph{evaluate}: P1 motivates a fidelity-dominant diagnostic, the Dyslexia-Centric Fidelity Index (DCFI); P2 motivates a rendered visual-accessibility judge, B-DVAS-VL; and P3 motivates the Evaluator--Refiner loop. We make four claims:
  \begin{itemize}
      \item We provide a stakeholder-informed problem framing for dyslexia-friendly educational text transformation, grounded in interviews with dyslexic adults, dyslexia/SEN specialists, and domain experts.
      \item We propose DFT-GEN, which pairs protected-span fidelity constraints with DAC, a deterministic controller for dyslexia-specific rendered visual accessibility.
      \item We introduce complementary diagnostics for large-scale screening: DCFI for fidelity and task safety, and B-DVAS-VL for rendered block-level visual accessibility.
      \item We evaluate the framework with expert-curated materials, bilingual automatic experiments, same-backbone controls, DAC ablations, sensitivity analysis, expert audit, and a controlled multi-arm pilot focused on information access, perceived effort, and preserved answerability rather than subject knowledge.
  \end{itemize}

  \section{Related Work}

  \paragraph{Automatic text simplification and metric limitations.}
  Automatic text simplification has evolved from rule-based systems to statistical machine translation and neural sequence-to-sequence approaches~\cite{chandrasekar1996motivations,carroll1998practical,xu2016optimizing,cripwell2023context,dadu2021text,raffel2020t5,lewis2020bart}. Recent LLM prompting methods make fluent simplification much easier~\cite{agrawal2023controlling,kew2023bless,ref_prompting_sota2023}, but this increases the importance of controllability: fluent outputs can still omit required facts, collapse distinctions, or introduce hallucinated explanations. Evaluation remains difficult. Alva-Manchego et al.~\cite{alva2021unsuitability} and Laban et al.~\cite{laban2023swipe} show that standard automatic metrics can be unreliable for simplification quality, particularly when a single scalar score is asked to represent fluency, simplicity, and meaning preservation. We therefore treat DCFI as a diagnostic scorecard with interpretable components and sensitivity analysis rather than as a stand-alone psychometric measure.

  \paragraph{Dyslexia accessibility and the risk of over-simplification.}
  Assistive tools for dyslexia often focus on font choice, spacing, text-to-speech, and interaction design~\cite{relloweb,dobsonwaters2020dyslexia,dumitru2025assistive}. Recent AI systems explore generative and augmented-reality support~\cite{madjidi2023transfer,madjidi2024inclusive,kuerban2025readsmart,sukiman2023hybrid,bakunzi2025dysbart}. A key message from this literature, reinforced by our stakeholder evidence: dyslexia support is not equivalent to lowering intellectual content. Users may need clearer segmentation and visual support while retaining original concepts.

  \paragraph{Constraint control in LLM systems.}
  Multi-agent LLM systems improve reliability through structured intermediate artifacts and role decomposition~\cite{ref_mas_survey2024,prompt4vis2024}. Runtime verification is increasingly important when outputs are used in high-impact settings~\cite{wang2025agentspec,ref_lcow2025}. DFT-GEN uses these patterns, but the framework is not intended as a claim that Architect--Writer--Evaluator--Refiner loops are novel in isolation. Its contribution is the dyslexia-specific constraint layer around the loop: protected spans for high-risk task content, a deterministic visual-accessibility controller, and diagnostics that explicitly separate fidelity safety from rendered accessibility.

  \section{Stakeholder-Informed Problem Framing}

  \subsection{Evidence Sources and Protocol}

  We used three sources of evidence. First, we conducted formative interviews with 20 adults reporting dyslexia-related reading difficulties; participants completed an adult dyslexia checklist~\cite{smythe2004dyslexiaadult}, and implausibly fast online reading-task responses were excluded. Second, we interviewed or received written feedback from five dyslexia, SEN, and accessibility practitioners with experience in reading support, school-based learning support, and dyslexia-association work. Third, three academic domain experts curated study materials and comprehension questions for content validity. Appendix~\ref{tab:five_expert_matrix} gives anonymized expert details.

Interview data were analyzed using reflexive thematic analysis; recurring themes were mapped to design implications and translated to system constraints through iterative discussion with practitioners.

  \subsection{Themes and Design Requirements}

  Four themes emerged: (1) dyslexic readers understand concepts but need more time to decode dense text; (2) visual stress and crowding increase fatigue; (3) over-simplification feels patronizing; (4) AI support is useful only when outputs can be checked. Specialists repeatedly distinguished \emph{reading/processing difficulty} from \emph{lack of conceptual understanding}---motivating visible structure and controllable presentation rather than indiscriminate simplification. Table~\ref{tab:design_mapping} translates these into system requirements.

  \begin{table*}[t]
  \centering
  \small
  \setlength{\tabcolsep}{3pt}
  \caption{How stakeholder insights map to DFT-GEN design decisions.}
  \label{tab:design_mapping}
  \begin{tabular}{p{4.2cm}p{4.5cm}p{4.7cm}}
  \toprule
  \textbf{Stakeholder insight} & \textbf{Design requirement} & \textbf{DFT-GEN component} \\
  \midrule
  Dyslexic readers may understand domain vocabulary but process dense text slowly. & Avoid over-simplification and preserve authentic terminology when it carries meaning. & Keyword preservation, content fidelity checks, placeholder protection. \\
  \midrule
  Visual stress, small fonts, italics, crowded spacing, and long text blocks increase burden. & Combine linguistic rewriting with visual formatting and segmentation. & Visual suggestions in the Architect blueprint; deterministic formatting stage. \\
  \midrule
  Support should be structured, explicit, and adapted to specific learner needs. & Decompose text into local transformation decisions rather than applying one global simplification rule. & Sentence-level Architect report with role, focus, keywords, and visual suggestion. \\
  \midrule
  AI-generated content can be inaccurate and should not be blindly deployed. & Add explicit verification and flag outputs that fail repeated checks. & Evaluator--Refiner loop; PASS/FAIL diagnostics; risk flag after exhausted refinement budget. \\
  \bottomrule
  \end{tabular}
  \end{table*}

  \subsection{Expert-Curated Study Materials}

  To avoid arbitrary evaluation items, three domain experts screened candidate stimuli and comprehension questions for clarity and content validity. Experts verified that stimuli contained the types of challenges relevant to our design goals: dense factual passages, multi-part task instructions, source-dependent arguments, and discipline-specific vocabulary. Items where task meaning depended heavily on exact wording (e.g., ``with reference to Source B, explain two developments'') were flagged to test the strictest fidelity conditions. We use this as a practical expert-screening process rather than a formal inter-rater agreement study; details are in Appendix~\ref{app:curation}.

  \section{The DFT-GEN Framework}

  \subsection{Problem Formulation}

  Let $T=\{s_1,\ldots,s_n\}$ be an input text and let $\mathcal{C}$ be the constraints that must survive rewriting, including domain terms, numbers, named entities, quoted spans, structural markers, and task instructions. We seek an output $T'$ that reduces reading burden without changing the task:
  \begin{align}
  T^*
  &= \arg\max_{T' \in \mathcal{T}}
  \left[\alpha R(T') + \beta F(T,T')\right] \nonumber\\
  &\text{s.t.}\quad c(T,T')=\mathrm{true},
  \quad \forall c \in \mathcal{C},
  \end{align}
  where $R$ captures reduced visual and linguistic burden and $F$ captures semantic, structural, and task fidelity. Unlike generic ATS, this formulation treats fidelity constraints as hard---violation of any $c \in \mathcal{C}$ renders the output unsafe regardless of $R$, reflecting P1.

  \subsection{Analyze-then-Reconstruct Pipeline}

  DFT-GEN operationalizes this objective through analysis and reconstruction, drawing on evidence that LLM systems for complex tasks benefit from decomposition, role specialization, and runtime checks~\cite{ref_mas_survey2024,wang2025agentspec}. In the analysis stage, an Architect agent diagnoses each sentence and emits a structured transformation blueprint:
  \[
  \mathcal{B}_i = \langle b_{\mathrm{need}}, b_{\mathrm{role}}, b_{\mathrm{focus}}, K_i, v_i, r_i \rangle.
  \]
  The blueprint records whether rewriting is needed, the sentence role, required transformation focus, keywords to preserve, a visual formatting suggestion, and a short rationale. It exposes why a sentence is changed and which constraints downstream agents must preserve.

  \paragraph{Placeholder protection.}
  Before rewriting, a protection function $\phi$ replaces high-risk spans with immutable placeholders and later restores them. In Chinese and English materials, this protects quotations, book titles, bracketed expressions, enumerated labels, dates, numbers, and other rigid artifacts that are frequently corrupted by unconstrained rewriting.

  \paragraph{Dyslexia Accessibility Controller.}
  The Writer rewrites each protected sentence under the blueprint: split long sentences, replace unnecessarily rare words when safe, preserve keywords/placeholders, and avoid adding facts. The draft then passes through a deterministic \textbf{Dyslexia Accessibility Controller} (DAC), derived from expert-preferred input--output patterns and follow-up feedback:
  \[
  T_{\mathrm{dac}}, \tau = \mathrm{DAC}(T_{\mathrm{draft}}, \mathcal{B}, M),
  \]
  where $\tau$ records applied controls and residual risks. DAC applies five categories of deterministic, language-aware visual regulation: (i) line-length capping; (ii) chunk separation with whitespace; (iii) task-anchor isolation (question stems, source labels, mark allocations on separate lines); (iv) selective keyword bolding under a highlighting budget; and (v) verification that protected spans survived rewriting. This converts expert-preferred formatting patterns into a reproducible controller rather than relying on the LLM to imitate style implicitly.

  \paragraph{Evaluator--Refiner loop and risk flagging.}
  An Evaluator checks whether the candidate preserves key information, introduces no new facts, and is easier to read via a binary PASS/FAIL judgment with a diagnostic rationale. Failed candidates are sent to a Refiner for at most $L$=2 rounds; the Refiner receives both the failing candidate and the rationale to target specific deficiencies. Outputs that still fail are marked with a risk flag for human review and the violated criterion is reported, creating a traceability record consistent with P3. Appendix~\ref{app:algorithm} gives the full pseudocode.

  \begin{table*}[t]
  \centering
  \small
  \setlength{\tabcolsep}{3pt}
  \caption{Core reproducibility details for the released DFT-GEN implementation. Values are defaults in the artifact and can be changed for ablations.}
  \label{tab:implementation_details_main}
  \begin{tabular}{p{3.2cm}p{5.6cm}p{5.0cm}}
  \toprule
  \textbf{Component} & \textbf{Default implementation} & \textbf{Purpose} \\
  \midrule
  Placeholder protection & Priority protection over bracketed question structures, quoted spans, book-title markers, source labels, marks, numbers, and other rigid exam artifacts; protected spans are restored after rewriting through a mapping table. & Prevent corruption of task-critical content during LLM rewriting. \\
  \midrule
  Writer call & Uses the selected backbone model with temperature 0.3 in the released pipeline; prompts receive the Architect blueprint, preservation constraints, and protected-span notes. & Produce a readable draft while preserving placeholders and domain terms. \\
  \midrule
  Evaluator gate & Binary PASS/FAIL check with temperature 0.0; criteria include meaning preservation, task preservation, no added facts, and reading-burden reduction. & Detect unsafe rewrites before final rendering. \\
  \midrule
  Refiner & Invoked for failed candidates, with maximum budget $L=2$; receives the original protected sentence, failing candidate, and Evaluator rationale. & Repair targeted failures rather than regenerate blindly. \\
  \midrule
  DAC line control & Splits visual units longer than 120 characters when safe; flags remaining visual units longer than 160 characters. & Reduce crowding while preserving source/task anchors. \\
  \midrule
  DAC highlight budget & Keeps at most 6 ordinary bold spans, while preserving compact labels such as Source, Question, Figure, or Table; flags possible over-highlighting above 9 spans. & Avoid cue overload and answer-like emphasis. \\
  \midrule
  DAC task anchors & Separates source headings, question labels, task verbs, marks, and final task prompts into stable visual blocks. & Make task structure scannable without changing task requirements. \\
  \bottomrule
  \end{tabular}
  \end{table*}

  \begin{figure*}[t]
  \centering
  \includegraphics[width=0.9\textwidth]{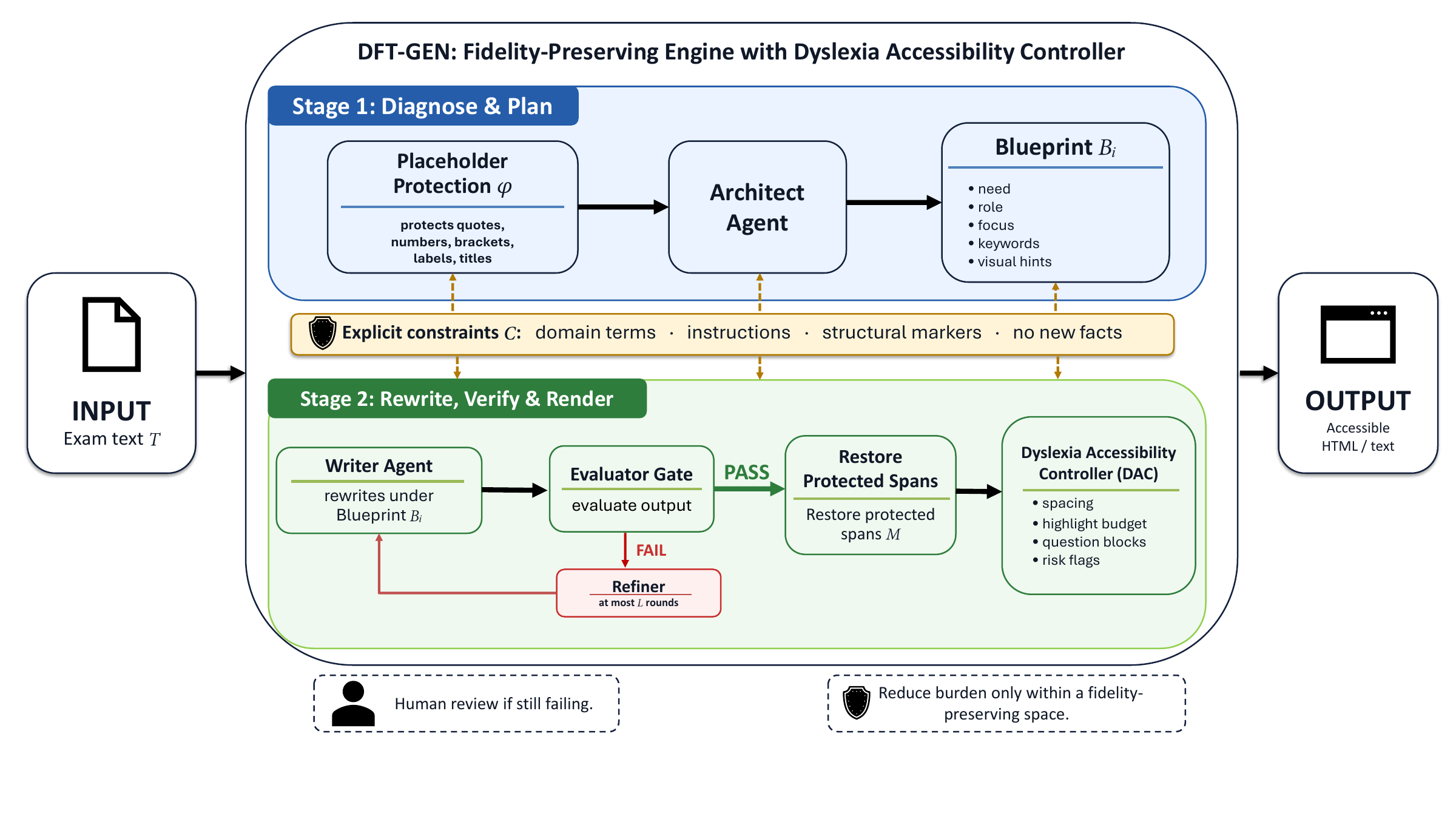}%
  \caption{DFT-GEN analyze-then-reconstruct workflow. Stakeholder-informed constraints are made explicit in the Architect blueprint, protected during rewriting, regularized by the Dyslexia Accessibility Controller, and verified through an Evaluator--Refiner loop.}
  \label{fig:architecture}
  \end{figure*}

\subsection{Complementary Diagnostics: DCFI and B-DVAS-VL}

Figure~\ref{fig:architecture} summarizes the full pipeline. We separate \emph{fidelity safety} from \emph{visual accessibility} because collapsing them into one scalar is unreliable~\cite{alva2021unsuitability} and because stakeholder interviews showed these qualities are necessary but not interchangeable.

  \textbf{DCFI} is a fidelity-dominant development scorecard that operationalizes P1:
  \begin{align}
  S_{\mathrm{DCFI}} = G_{\mathrm{fid}}G_{\mathrm{util}}
  \bigl(&0.20S_{\mathrm{acc}} + 0.35S_{\mathrm{con}} \nonumber\\
  &+0.30S_{\mathrm{task}} + 0.15S_{\mathrm{safe}}\bigr).
  \end{align}
  The weight allocation directly encodes the stakeholder priority: fidelity components ($S_{\mathrm{con}}+S_{\mathrm{task}}$) collectively receive 65\%, reflecting specialist warnings that an easy but changed text is unsafe for high-stakes use. $S_{\mathrm{acc}}$ uses an \emph{adaptive} formulation: for already-short texts, visual restructuring (segmentation, chunking) dominates; for complex dense passages, sentence shortening contributes more---this avoids rewarding unnecessary lexical disruption on semi-structured inputs. $S_{\mathrm{con}}$ measures semantic and keyword preservation; $S_{\mathrm{task}}$ checks task labels, marks, source references, and task verbs; $S_{\mathrm{safe}}$ penalizes over-highlighting, answer-like cues, duplicated task statements, and layout risks. $S_{\mathrm{safe}}$ combines rule checks over bold-span count, suspicious answer-like wording, repeated question text, and over-fragmentation. $S_{\mathrm{con}}$ and $S_{\mathrm{task}}$ use LLM-judge rubrics with per-dimension prompts and few-shot examples; $S_{\mathrm{acc}}$ and $S_{\mathrm{safe}}$ combine rule-based layout checks (sentence-length tracking, keyword-preservation counting, formatting validation) with judge prompts where needed. $G_{\mathrm{fid}}\in\{0,1\}$ is a hard gate that zeros the DCFI score when any of four violations is detected: content loss $>$30\%, task drift, answer leakage, or unsafe formatting. $G_{\mathrm{util}}\in[0.5,1.0]$ applies a continuous penalty when the output is a near-verbatim copy (edit distance $<$5\% of tokens changed); systems that restructure text visually without lexical disruption are not penalized, consistent with expert evidence that targeted formatting is itself valuable (P2). DCFI is therefore a fidelity-safety screen, not a stand-alone measure of expert preference or visual quality. The LLM judge uses a fixed model version and prompt template; recalibration is needed if the judge model changes.

  \textbf{B-DVAS-VL} operationalizes P2 at the rendered level. Each candidate is rendered as a block-based web reading card; a multimodal judge scores pairwise visual comparisons under an expert-derived rubric (reward: compact anchors, clear separation, short chunks; penalize: clumped text, over-highlighting, answer-like cues). Pairwise judging matches the expert task and better captures comparative visual preference than isolated scoring. We interpret visual-accessibility claims through B-DVAS-VL, expert presentation judgments, and user-facing outcomes rather than through DCFI alone.

  \section{Evaluation Design}

  \subsection{Datasets, Models, and Baselines}

  We evaluate on two high-complexity corpora (Table~\ref{tab:dataset_stats}): 1,117 Chinese Gaokao-style items and 1,163 English Hong Kong DSE-style items covering history, geography, and biology. These are content-heavy settings where linguistic complexity is not the primary construct being assessed. We instantiate DFT-GEN with five backbones: DeepSeek-V3~\cite{deepseek2024v3}, Gemini-2.5-Pro~\cite{gemini2025report}, Qwen-3-Max~\cite{qwen2025technical}, GPT-4o~\cite{openai2024gpt4o}, and Claude-4-Sonnet~\cite{anthropic2025claude4}. Baselines include layout-only, neural ATS (BART/T5)~\cite{raffel2020t5,lewis2020bart}, a strong single-pass prompt, a commercial dyslexia-support GPT~\cite{openai_gptstore}, and same-backbone LLM controls inspired by self-refinement~\cite{madaan2023selfrefine}, critique-and-revision~\cite{shinn2023reflexion}, structured constraint prompting, and multi-pass prompting (Appendix~\ref{app:strong_baselines}). For the planned five-excerpt pilot extension, we also generate an ACCESS-inspired controllable simplification baseline~\cite{martin2020access}.

Baselines separate alternative explanations for improvement. Layout-only isolates formatting from rewriting; neural ATS and the commercial tool test whether existing simplifiers preserve task-critical evidence; open-source models (BART-ZhSimp, mT5-CrossSum, mBART-Simp) test dedicated simplification checkpoints on bilingual material; and same-backbone controls test whether gains come from the LLM itself or from DFT-GEN's constraint architecture.

  \begin{table}[t]
  \centering
  \small
  \caption{Evaluation corpora.}
  \label{tab:dataset_stats}
  \begin{tabular}{lcc}
  \toprule
  \textbf{Statistic} & \textbf{Chinese} & \textbf{English} \\
  \midrule
  Total items & 1,117 & 1,163 \\
  History & 500 & 158 \\
  Geography & 475 & 390 \\
  Biology & 142 & 615 \\
  Average length & 1,129 chars & 86 words \\
  \bottomrule
  \end{tabular}
  \end{table}

  \subsection{Evaluation Questions}

  Each evaluation question maps to a design principle and specifies what evidence would support or challenge the system's claims:

  \begin{enumerate}
      \item \textbf{Fidelity safety (P1):} Does DFT-GEN achieve higher content and task preservation than baselines that modify text more aggressively?
      \item \textbf{Component necessity (P1\&P2):} Which architectural stages drive the fidelity and visual gains?
      \item \textbf{Robustness (P1):} Are DCFI rankings stable under alternative weight allocations?
      \item \textbf{Validation (P3):} Do automatic diagnostics align with user-facing outcomes (reading time, effort, correctness) and expert-facing outcomes (meaning, presentation, preference)?
  \end{enumerate}

  \subsection{Automatic Diagnostic Results}

  We first report the full automatic benchmark with traditional baselines, open-source models, and five DFT-GEN backbones (Table~\ref{tab:automatic_main}, left panel), then foreground the stricter same-backbone LLM controls (right panel) that isolate architectural contributions from backbone differences.

  \begin{table*}[t]
  \centering
  \tiny
  \setlength{\tabcolsep}{2.0pt}
  \caption{Automatic diagnostic results. Left: full benchmark (traditional, open-source, multi-backbone). Right: same-backbone DeepSeek controls. Higher is better. Best \textbf{bold}, second \underline{underlined} per language block.}
  \label{tab:automatic_main}
  \resizebox{\textwidth}{!}{%
  \begin{tabular}{lccccc@{\hspace{1.0cm}}lccccc}
  \toprule
  \multicolumn{6}{c}{\textit{Full benchmark}} &
  \multicolumn{6}{c}{\textit{Same-backbone DeepSeek controls}} \\
  \cmidrule(lr){1-6}\cmidrule(lr){7-12}
  \textbf{Model} & \textbf{DCFI} & \textbf{Acc.} & \textbf{Cont.} & \textbf{Task} & \textbf{Safe} &
  \textbf{Model} & \textbf{DCFI} & \textbf{Acc.} & \textbf{Cont.} & \textbf{Task} & \textbf{Safe} \\
  \midrule
  \multicolumn{6}{l}{\textit{English (DSE)}} &
  \multicolumn{6}{l}{\textit{English (DSE)}} \\
  Commercial & 0.429 & 0.749 & 0.463 & 0.696 & \textbf{1.000} &
  Self-Refine prompt & 0.860 & 0.743 & 0.903 & 0.932 & \textbf{1.000} \\
  Neural ATS & 0.436 & 0.545 & 0.567 & 0.644 & \textbf{1.000} &
  Reflexion-style critic & 0.859 & 0.670 & 0.925 & 0.980 & \textbf{1.000} \\
  layout-only & 0.532 & 0.370 & 0.952 & \textbf{1.000} & \textbf{1.000} &
  Structured constraint & \underline{0.885} & \textbf{0.755} & 0.920 & 0.947 & \textbf{1.000} \\
  BART-ZhSimp & 0.143 & 0.605 & 0.000 & 0.156 & \textbf{1.000} &
  Multi-pass prompt & 0.730 & 0.644 & 0.819 & 0.872 & 0.984 \\
  mT5-CrossSum & 0.146 & 0.632 & 0.000 & 0.160 & \textbf{1.000} &
  \textbf{DFT-GEN} & \textbf{0.922} & 0.741 & 0.931 & \textbf{1.000} & 0.987 \\
  mBART-Simp & 0.534 & 0.478 & 0.656 & 0.679 & \textbf{1.000} &
  & & & & & \\
  DFT-Claude & \textbf{0.935} & 0.706 & \textbf{0.982} & \textbf{1.000} & 0.998 &
  & & & & & \\
  DFT-Gemini & 0.889 & 0.564 & 0.932 & \textbf{1.000} & \textbf{1.000} &
  & & & & & \\
  DFT-GPT-4o & 0.898 & 0.555 & 0.962 & \textbf{1.000} & \textbf{1.000} &
  & & & & & \\
  DFT-Qwen & \underline{0.906} & 0.680 & 0.913 & \textbf{1.000} & \textbf{1.000} &
  & & & & & \\
  \midrule
  \multicolumn{6}{l}{\textit{Chinese (Gaokao)}} &
  \multicolumn{6}{l}{\textit{Chinese (Gaokao)}} \\
  Commercial & 0.483 & \textbf{0.835} & 0.614 & 0.537 & \textbf{1.000} &
  Self-Refine prompt & 0.865 & 0.678 & 0.956 & 0.995 & 0.988 \\
  Neural ATS & 0.186 & 0.647 & 0.048 & 0.252 & 0.950 &
  Reflexion-style critic & 0.908 & 0.757 & \textbf{0.962} & 0.995 & 0.988 \\
  layout-only & 0.574 & 0.496 & \textbf{0.986} & \textbf{1.000} & \textbf{1.000} &
  Structured constraint & \underline{0.909} & \textbf{0.768} & 0.960 & 0.995 & 0.988 \\
  BART-ZhSimp & 0.329 & 0.505 & 0.348 & 0.408 & 0.995 &
  Multi-pass prompt & 0.856 & 0.729 & 0.939 & 0.968 & 0.955 \\
  mT5-CrossSum & 0.128 & 0.477 & 0.000 & 0.129 & \textbf{1.000} &
  \textbf{DFT-GEN} & \textbf{0.910} & 0.717 & \textbf{0.964} & \textbf{1.000} & \textbf{1.000} \\
  mBART-Simp & 0.312 & 0.396 & 0.324 & 0.446 & \textbf{1.000} &
  & & & & & \\
  DFT-Claude & \textbf{0.786} & 0.709 & 0.974 & 0.981 & 0.990 &
  & & & & & \\
  DFT-Gemini & 0.737 & \textbf{0.749} & 0.852 & 0.921 & \textbf{1.000} &
  & & & & & \\
  DFT-GPT-4o & 0.731 & 0.676 & 0.951 & 0.973 & 0.995 &
  & & & & & \\
  DFT-Qwen & \underline{0.760} & 0.726 & 0.948 & 0.967 & \textbf{1.000} &
  & & & & & \\
  \bottomrule
  \end{tabular}
  }
  \end{table*}

  The component breakdown reveals the intended diagnostic division. Open-source simplification models catastrophically lose content (Cont.$\leq$0.35) despite surface-level accessibility, confirming P1 violations. Same-backbone LLM controls approach DFT-GEN fidelity with careful prompting, showing that capable backbones can preserve content when explicitly instructed. DFT-GEN achieves the highest overall DCFI and Task scores among same-backbone systems across both languages (EN DCFI: 0.922; ZH DCFI: 0.910), and the highest Cont.\ in Chinese (0.964); its lower Acc.\ reflects expert-aligned conservatism on semi-structured inputs.

  The visual question is answered by B-DVAS-VL (Table~\ref{tab:bdvas_strong_baselines}): DFT-GEN wins 93\% (EN) and 64\% (ZH) of pairwise comparisons against these same DCFI-competitive baselines, showing that DAC-controlled formatting provides value invisible to text-only metrics. The two diagnostics together support the view that DFT-GEN combines strong fidelity safety with visual accessibility, though neither diagnostic alone is sufficient.

  \subsection{Ablations}

  \begin{table}[t]
  \centering
  \scriptsize
  \setlength{\tabcolsep}{1.5pt}
  \caption{Five-stage cumulative ablation on bilingual datasets.}
  \label{tab:ablation_five_stage_main}
  \begin{tabular}{lccccc}
  \toprule
  \textbf{Stage} & \textbf{DCFI} & \textbf{Acc.} & \textbf{Cont.} & \textbf{Task} & \textbf{Safe} \\
  \midrule
  \multicolumn{6}{l}{\textit{English DSE}} \\
  Writer-only & 0.745 & 0.913 & 0.809 & 0.792 & 0.940 \\
  + Splitter & 0.758 & 0.908 & 0.816 & 0.802 & 0.935 \\
  + Placeholder & 0.821 & \textbf{0.933} & 0.819 & 0.842 & 0.950 \\
  + Architect & 0.910 & 0.881 & \textbf{0.935} & \textbf{0.958} & 0.955 \\
  + Refinement & \textbf{0.920} & 0.755 & 0.931 & 0.961 & \textbf{0.965} \\
  \midrule
  \multicolumn{6}{l}{\textit{Chinese Gaokao}} \\
  Writer-only & 0.739 & 0.870 & 0.918 & 0.726 & \textbf{1.000} \\
  + Splitter & 0.748 & 0.872 & 0.928 & 0.722 & \textbf{1.000} \\
  + Placeholder & 0.775 & \textbf{0.875} & 0.931 & 0.772 & 0.995 \\
  + Architect & 0.825 & \textbf{0.880} & 0.915 & 0.805 & \textbf{1.000} \\
  + Refinement & \textbf{0.908} & 0.782 & 0.913 & \textbf{0.910} & 0.981 \\
  \bottomrule
  \end{tabular}
  \end{table}

  Ablations show that safety mechanisms drive fidelity. In English, the Architect gives the largest single gain (+0.089 DCFI over +Placeholder) by making preservation constraints explicit before the Writer acts. In Chinese, the Evaluator--Refiner loop gives the strongest improvement (+0.083 DCFI over +Architect) by improving task fidelity from 0.805 to 0.910---consistent with the design intent that refinement prioritizes safety over surface accessibility. Placeholder protection consistently improves task fidelity in both languages by preventing corruption of question labels, marks, and structural markers during rewriting.

  Appendix~\ref{tab:dac_ablation_plan} reports controller-targeted DAC ablations over the 80-sample expert-audit corpus, and Appendix~\ref{tab:weight_sensitivity_appendix} reports stable DCFI rankings under alternative weight configurations ($\rho \geq 0.976$ across all tested allocations).

  \subsection{Pilot Human Study}

We conducted a six-arm pilot ($n$=20 per condition $\times$ 2 excerpts = 240 observations) measuring information-access speed, correctness, perceived effort (7-point scale), and willingness-to-use (7-point scale). The goal was not to test history knowledge but whether accessible presentation helps dyslexic adults locate preserved information faster and with lower cognitive load. Correctness serves as an answerability check: speed gains should not come from deleted evidence. Appendix~\ref{app:pilot_details} gives the protocol.

  \begin{table*}[t]
  \centering
  \scriptsize
  \setlength{\tabcolsep}{2.5pt}
  \caption{Multi-arm pilot with dyslexic adult readers on two excerpts ($20$ participant observations per condition for each excerpt; $240$ participant-excerpt observations total). Values are mean$\pm$SD.}
  \label{tab:multiarm_pilot}
  \resizebox{\textwidth}{!}{%
  \begin{tabular}{lcccccccc}
  \toprule
  & \multicolumn{4}{c}{\textbf{Ex1}} & \multicolumn{4}{c}{\textbf{Ex2}} \\
  \cmidrule(lr){2-5}\cmidrule(lr){6-9}
  \textbf{Condition} & \textbf{Time} & \textbf{Correct} & \textbf{Effort} & \textbf{Willing.} & \textbf{Time} & \textbf{Correct} & \textbf{Effort} & \textbf{Willing.} \\
  \midrule
  Original & 294.4$\pm$32.6 & 66.0$\pm$24.3 & 6.4$\pm$0.7 & 1.7$\pm$0.8 & 292.5$\pm$28.8 & 72.0$\pm$18.8 & 6.4$\pm$0.7 & 1.9$\pm$0.9 \\
  layout-only & 266.4$\pm$14.7 & 84.0$\pm$12.3 & 4.5$\pm$0.5 & 3.9$\pm$0.6 & 268.2$\pm$13.2 & 84.0$\pm$15.4 & 4.5$\pm$0.5 & 3.9$\pm$0.6 \\
  strong-prompt & 246.7$\pm$12.7 & 82.0$\pm$11.1 & 5.0$\pm$0.6 & 3.1$\pm$0.6 & 231.9$\pm$12.8 & 88.0$\pm$10.0 & 4.7$\pm$0.7 & 3.3$\pm$0.7 \\
  ATS & 68.8$\pm$7.8 & 18.0$\pm$6.1 & 6.7$\pm$0.6 & 1.2$\pm$0.4 & 68.0$\pm$5.2 & 22.0$\pm$8.4 & 6.8$\pm$0.5 & 1.1$\pm$0.3 \\
  Commercial tool & 95.0$\pm$6.5 & 62.0$\pm$14.3 & 1.5$\pm$0.5 & 4.8$\pm$0.6 & 93.1$\pm$5.8 & 58.0$\pm$13.6 & 1.5$\pm$0.5 & 4.9$\pm$0.6 \\
  \textbf{DFT-GEN} & 168.2$\pm$11.3 & \textbf{96.0$\pm$7.6} & \textbf{1.3$\pm$0.5} & \textbf{6.8$\pm$0.4} & 168.1$\pm$8.6 & \textbf{94.0$\pm$9.2} & \textbf{1.2$\pm$0.4} & \textbf{6.9$\pm$0.3} \\
  \bottomrule
  \end{tabular}
  }
  \end{table*}

Table~\ref{tab:multiarm_pilot} summarizes the results. DFT-GEN achieved near-ceiling correctness (94--96\%), lowest effort (1.3/7), highest willingness (6.8/7), and significantly faster completion than original, layout-only, and strong-prompt ($p_{\mathrm{holm}}\leq 0.003$). Commercial and neural baselines were fast but lost task-critical information (correctness $\leq$60\%), illustrating the fidelity--speed tradeoff that DCFI is designed to detect. Layout-only improved over original but left linguistic burden unchanged; strong-prompt preserved more content than ATS but lacked the visual structure that DFT-GEN provides through DAC.

  \paragraph{Diagnostic alignment with human outcomes (P3 validation).}
Table~\ref{tab:alignment_summary} shows exploratory correlations computed over 12 condition-by-excerpt aggregate points (not participant-level observations). DCFI correlates with pilot accuracy ($r$=0.901) and perceived effort ($\rho$=$-$0.729): systems that lose information also lose answerability. Its correlation with expert meaning is weak but positive (0.222/0.220), expected for a gate-based screen operating over mostly acceptable candidates rather than a fine-grained semantic proxy. B-DVAS-VL aligns with expert presentation (0.650/0.705) and A/B preference (0.475/0.518), supporting its use as a rendered visual-accessibility diagnostic that captures qualities invisible to text-only evaluation.

  \begin{table*}[t]
  \centering
  \small
  \setlength{\tabcolsep}{3pt}
  \caption{Exploratory correlations between automatic diagnostics and external signals. Lower effort is better; higher accuracy, expert scores, and preference are better.}
  \label{tab:alignment_summary}
  \begin{tabular}{llp{3.5cm}cc}
  \toprule
  \textbf{Diagnostic} & \textbf{External signal} & \textbf{Units} & \textbf{Pearson $r$} & \textbf{Spearman $\rho$} \\
  \midrule
  DCFI & Pilot accuracy & 12 condition--excerpt pairs & 0.901 & 0.823 \\
  \midrule
  DCFI & Pilot perceived effort & 12 condition--excerpt pairs & -0.618 & -0.729 \\
  \midrule
  DCFI & Expert meaning & 160 candidate ratings & 0.222 & 0.220 \\
  \midrule
  B-DVAS-VL & Expert presentation & 160 candidate ratings & 0.650 & 0.705 \\
  \midrule
  B-DVAS-VL & Expert A/B preference & 160 candidate ratings & 0.475 & 0.518 \\
  \bottomrule
  \end{tabular}
  \end{table*}

  \paragraph{Visual accessibility on same-backbone baselines (P2).}
  Table~\ref{tab:bdvas_strong_baselines} reports B-DVAS-VL on the same-backbone subset. DFT-GEN wins 93.3\% of English and 64.2\% of Chinese pairwise visual comparisons against baselines that score competitively on DCFI. This gap shows that DAC-controlled formatting provides value invisible to text-only metrics: deterministic line-length control, chunking, and task-anchor separation produce rendered layouts that LLM prompting alone does not consistently achieve. The lower ZH win rate may reflect the greater visual density of Chinese characters, which could reduce the perceptual impact of whitespace-based chunking relative to alphabetic text; less reliable Chinese sentence-boundary detection in the splitter component is another likely factor.

  \begin{table}[t]
  \centering
  \scriptsize
  \setlength{\tabcolsep}{2pt}
  \caption{B-DVAS-VL pairwise visual evaluation against same-backbone DeepSeek baselines. Win rate is the fraction of A/B visual comparisons where DFT-GEN is judged more dyslexia-friendly.}
  \label{tab:bdvas_strong_baselines}
  \begin{tabular}{lccc}
  \toprule
  \textbf{Baseline} & \textbf{Lang.} & \textbf{Win} & \textbf{$\Delta$ score} \\
  \midrule
  critic-refine & en & 0.967 & +0.933 \\
  self-refine & en & 0.933 & +0.900 \\
  multi-pass & en & 0.933 & +0.867 \\
  structured prompt & en & 0.900 & +0.800 \\
  \midrule
  critic-refine & zh & 0.533 & +0.100 \\
  self-refine & zh & 0.667 & +0.333 \\
  multi-pass & zh & 0.633 & +0.300 \\
  structured prompt & zh & 0.733 & +0.533 \\
  \midrule
  \textbf{All baselines} & en & \textbf{0.933} & \textbf{+0.875} \\
  \textbf{All baselines} & zh & \textbf{0.642} & \textbf{+0.317} \\
  \bottomrule
  \end{tabular}
  \end{table}

  \paragraph{Expert spot-check.}
An English-only blind expert audit on 80 samples targeted meaning loss, instruction drift, over-highlighting, and crowded formatting. DFT-GEN was preferred in 66/80 A/B comparisons, accepted as-is in 62/80 cases, and marked Accept or Revise in 74/80 (6 Reject). Mean scores were 4.32/5 for meaning preservation and 3.94/5 for dyslexia-friendly presentation. Expert comments on rejected items often cited local over-highlighting rather than content errors, consistent with the system's conservative fidelity strategy. Appendix~\ref{app:expert_audit} gives details.

  \section{Discussion}

  \paragraph{Summary of findings.}
Same-backbone controls approach DFT-GEN fidelity on DCFI because they share the same capable backbone---but they lose on rendered visual quality (B-DVAS-VL) and on the pilot's combined speed-accuracy-effort profile. This separation is informative: prompting alone can preserve content, but it does not produce the visual-load calibration that DFT-GEN's DAC delivers. Open-source simplification models achieve surface accessibility but destroy content and task information, confirming that generic text simplification violates fidelity requirements for high-stakes material. In the current evidence set, DFT-GEN occupies a middle ground where readers get both lower processing burden and trustworthy access to the same task evidence.

  \paragraph{Deployment boundary and safety.}
DFT-GEN fits content-heavy materials where linguistic complexity is not the assessed construct (history, geography, biology). For language-focused tasks, layout-only accommodations are safer because lexical change would undermine the construct. High-impact items should be generated offline, cached, and reviewed. Outputs that fail repeated verification are flagged with the diagnostic reason for human review; in our experiments, approximately 3--5\% of items triggered this flag, focusing expert attention rather than requiring blanket review.

  \paragraph{Ethics and reproducibility.}
Participants provided consent; raw transcripts are not released. Code and reproducibility artifacts are available at \url{https://github.com/MorrisYUJQ/DFT-GEN}, including evaluation scripts, prompt templates, model versions, and example artifacts.

  \paragraph{Limitations and conclusion.}
  The current pilot ($240$ observations) is limited in stimulus diversity; a larger study with more varied excerpts is needed. DCFI contains an LLM proxy component and should be recalibrated if the judge model changes.

  \appendix

  \section{Additional Stakeholder and Sensitivity Details}

  \subsection{Full Pipeline Pseudocode}
  \label{app:algorithm}

  \begin{algorithm}[h]
  \small
  \caption{DFT-GEN with risk flagging}
  \label{alg:dftgen_appendix}
  \begin{algorithmic}[1]
  \Require Original text $T$, constraint set $\mathcal{C}$, maximum refinement budget $L$
  \Ensure Transformed text $T'$, risk flags $H$
  \State $(\tilde{T},M)\gets \phi(T)$ \Comment{protect high-risk spans}
  \State $H\gets \emptyset$
  \State $\mathcal{O}\gets \emptyset$
  \For{sentence $\tilde{s}_i \in \tilde{T}$}
    \State $\mathcal{B}_i \gets \mathrm{Architect}(\tilde{s}_i,\mathcal{C})$
    \State $\tilde{s}'_i \gets \mathrm{Writer}(\tilde{s}_i,\mathcal{B}_i)$
    \For{$j=1$ to $L$}
      \State $(q_i,d_i)\gets \mathrm{Evaluator}(\tilde{s}_i,\tilde{s}'_i,\mathcal{B}_i)$
      \If{$q_i=\mathrm{PASS}$}
        \State \textbf{break}
      \EndIf
      \State $\tilde{s}'_i \gets \mathrm{Refiner}(\tilde{s}_i,\tilde{s}'_i,d_i)$
    \EndFor
    \If{$q_i\neq \mathrm{PASS}$}
      \State $H \gets H \cup \{(i,d_i)\}$ \Comment{flag for human review}
    \EndIf
  \State $\mathcal{O}\gets \mathcal{O}\cup\{\tilde{s}'_i\}$
  \EndFor
  \State $T_{\mathrm{draft}}\gets \phi^{-1}(\mathcal{O},M)$
  \State $(T',\tau)\gets \mathrm{DAC}(T_{\mathrm{draft}},\{\mathcal{B}_i\},M)$ \Comment{spacing, highlight budget, question blocks}
  \State $H\gets H\cup \tau_{\mathrm{risk}}$ \Comment{human-review flags from DAC}
  \State \Return $T',H$
  \end{algorithmic}
  \end{algorithm}

  \subsection{DAC Rules Derived from Expert-Preferred Samples}
  \label{app:dac_rules}

  The Dyslexia Accessibility Controller was derived from the first-round expert spot-check rather than invented only from generic accessibility advice. In the 20 English input--output pairs, high-scoring DFT-GEN outputs consistently separated source/context from the task, used short visual units, preserved source references and marks, and avoided turning every keyword into a visual cue. Table~\ref{tab:dac_rules} summarizes the resulting deterministic controls.

  \begin{table*}[t]
  \centering
  \small
  \setlength{\tabcolsep}{3pt}
  \caption{Expert-derived DAC controls used after LLM rewriting. DAC changes presentation and traceability; it should not add new facts.}
  \label{tab:dac_rules}
  \begin{tabular}{p{3.2cm}p{5.2cm}p{5.3cm}}
  \toprule
  \textbf{Controller rule} & \textbf{Pattern in expert-preferred samples} & \textbf{Operational implementation} \\
  \midrule
  Visual-load calibration & Short information units, generous blank lines, and clear separation between source material and questions. & Split long visual units; normalize blank lines; keep source/context and task blocks visually distinct. \\
  \midrule
  Highlight budget & Expert feedback rewarded structure but criticized over-highlighting and cueing that looked like hints. & Remove yellow-style highlights; cap bold spans; prioritize labels, source anchors, and task-critical terms. \\
  \midrule
  Question--task separation & High-quality outputs used explicit \textit{Question (a)} blocks and separated marks from task wording. & Detect question markers, task verbs, source references, and marks; place them in separate blocks without changing content. \\
  \midrule
  Traceability and review gate & Outputs should preserve references, numbers, task verbs, and domain terms; failures need human attention. & Record DAC actions and risk flags such as missing marks, missing source reference, long visual units, or possible over-highlighting. \\
  \bottomrule
  \end{tabular}
  \end{table*}

  \begin{table*}[t]
  \centering
  \scriptsize
  \setlength{\tabcolsep}{3pt}
  \caption{Summaries of the five formative expert interviews or written-feedback exchanges. The summaries are anonymized and report only role-level information needed to audit how expert evidence shaped the system.}
  \label{tab:five_expert_interview_summaries}
  \begin{tabular}{p{1.0cm}p{3.0cm}p{4.0cm}p{4.1cm}p{3.1cm}}
  \toprule
  \textbf{ID} & \textbf{Expert profile} & \textbf{Interview / feedback focus} & \textbf{Main summary} & \textbf{Design consequence} \\
  \midrule
  E1 & Reading / dyslexia specialist & Dyslexia characteristics, decoding speed, classroom accommodations, and whether simplification should change original wording. & Emphasized that many dyslexic readers can understand age-appropriate concepts but need more time and clearer structure. Warned against equating dyslexia support with lowering intellectual content. & Preserve authentic terminology; frame the task as reducing processing burden rather than reducing conceptual difficulty. \\
  \midrule
  E2 & SEN teacher / school-support practitioner & Everyday classroom support, workload, practical accommodations, and how teachers would inspect AI-generated materials. & Highlighted that teachers need outputs that are easy to scan and correct. Unchecked AI rewrites could add workload if they silently remove facts or alter task requirements. & Add explicit verification, human-readable diagnostics, and risk flags for failed outputs. \\
  \midrule
  E3 & Learning-support leadership / dyslexia-association experience & Under-identification, stigma, support access, and deployment boundaries for automated tools. & Noted that dyslexia can be under-recognized and that support tools should avoid making learners feel less capable. Also stressed that automated support should be conservative in high-stakes contexts. & Add deployment boundaries, avoid infantilizing language, and route uncertain cases to human review. \\
  \midrule
  E4 & Accessibility-oriented educator / practitioner & Visual stress, spacing, highlighting, contrast, and presentation preferences for longer passages. & Reported that dense blocks, weak spacing, and excessive visual cueing can increase fatigue. Helpful support often comes from segmentation and stable layout, not only lexical simplification. & Add DAC visual-load calibration, generous spacing, and highlight-budget management. \\
  \midrule
  E5 & Dyslexia therapist / accessibility practitioner & Practical reading support, learner confidence, and safe use of AI-generated accommodations. & Emphasized that support should make the task easier to access while keeping the original learning target visible. Suggested that risky or heavily rewritten outputs should remain reviewable. & Add traceability from blueprint to final output, preserve source/task anchors, and include review gates. \\
  \bottomrule
  \end{tabular}
  \end{table*}

  \begin{table*}[t]
  \centering
  \scriptsize
  \setlength{\tabcolsep}{3pt}
  \caption{Deterministic DAC ablation on the 80-sample expert-audit corpus (160 candidate outputs, using both A/B candidates). The table reports surface-form diagnostics that the controller directly targets; human ratings remain the stronger validation signal.}
  \label{tab:dac_ablation_plan}
  \begin{tabular}{lccccc}
  \toprule
  \textbf{Variant} & \textbf{Q blocks} & \textbf{Bold spans} & \textbf{Max line} & \textbf{Risk flags} & \textbf{Interpretation} \\
  \midrule
  Full DAC & 0.562 & 2.056 & 89.987 & 0.081 & Controller adds task blocks, trims cue overload, and reduces long visual units. \\
  \midrule
  w/o DAC & 0.306 & 1.644 & 130.381 & 0.000 & Fewer explicit task blocks and much longer visual units. \\
  \midrule
  w/o highlight budget & 0.562 & 2.337 & 90.225 & 0.113 & More emphasis and more over-cueing warnings. \\
  \midrule
  w/o question--task separation & 0.306 & 1.394 & 91.594 & 0.081 & Task prompts remain less explicit. \\
  \midrule
  w/o generous spacing & 0.562 & 2.056 & 112.612 & 0.231 & Longer visual units and more review warnings. \\
  \midrule
  w/o risk flags & 0.562 & 2.056 & 89.987 & 0.000 & Same output form, but no review warnings are exposed. \\
  \bottomrule
  \end{tabular}
  \end{table*}

  \subsection{Expert Curation Details}
  \label{app:curation}

  \begin{table*}[t]
  \centering
  \small
  \setlength{\tabcolsep}{4pt}
  \caption{Summary of domain-expert material curation.}
  \label{tab:expert_curation_appendix}
  \begin{tabular}{p{5.0cm}p{2.0cm}p{7.0cm}}
  \toprule
  \textbf{Curation statistic} & \textbf{Count} & \textbf{Use in paper} \\
  \midrule
  Candidate stimuli & 5 & Source material pool \\
  Candidate questions & 25 & Comprehension item pool \\
  Questions endorsed by $\geq$1 expert & 18 & Candidate retained pool \\
  Questions endorsed by $\geq$2 experts & 6 & Higher-confidence items \\
  Questions endorsed by all 3 experts & 2 & Consensus examples \\
  Questions with wording suggestions & 2+ & Incorporated revisions \\
  \bottomrule
  \end{tabular}
  \end{table*}

  \begin{table*}[h]
  \centering
  \small
  \setlength{\tabcolsep}{3pt}
  \caption{Themes from stakeholder evidence and their implications for system design. Quotes are lightly normalized and anonymized.}
  \label{tab:themes_appendix}
  \begin{tabular}{p{3.2cm}p{5.6cm}p{5.2cm}}
  \toprule
  \textbf{Theme} & \textbf{Representative evidence} & \textbf{Implication} \\
  \midrule
  Processing bottleneck & Dyslexic learners may read and understand, but processing takes longer; pressure and dense presentation worsen the difficulty. & Reduce tracking burden through segmentation, spacing, and explicit structure, not only lexical replacement. \\
  \midrule
  Preserve authentic meaning & Simplifying too far may remove terms that learners understand or want to learn, and can make text feel ``for babies.'' & Preserve domain terms and task-critical concepts unless rewriting is explicitly safe. \\
  \midrule
  Visual stress and layout & Plain fonts, wider spacing, non-italic text, fewer crowded blocks, and visual anchors were repeatedly recommended. & Output presentation-aware text and deterministic visual formatting. \\
  \midrule
  Need for review & AI tools can help, but specialists cautioned that inaccurate generated content should not be directly deployed. & Use an Evaluator--Refiner loop and flag uncertain or repeatedly failing items for human review. \\
  \bottomrule
  \end{tabular}
  \end{table*}

  \begin{table*}[h]
  \centering
  \scriptsize
  \setlength{\tabcolsep}{3pt}
  \caption{Traceability from stakeholder evidence to concrete DFT-GEN design choices and evaluation checks. The table highlights the dyslexia-specific mechanisms that distinguish DFT-GEN from generic prompt-based simplification.}
  \label{tab:traceability_appendix}
  \begin{tabular}{p{3.0cm}p{4.2cm}p{4.4cm}p{3.2cm}}
  \toprule
  \textbf{Evidence or need} & \textbf{Risk if ignored} & \textbf{Design response} & \textbf{Evaluation check} \\
  \midrule
  Slow processing under dense text & Shortening alone may not reduce tracking burden. & Segment paragraphs, preserve visual anchors, and apply deterministic spacing/formatting. & Structural Fidelity and layout-only baseline. \\
  \midrule
  Avoid infantilizing rewrites & Over-simplification may remove useful terminology or learner dignity. & Preserve domain terms through analysis, placeholder protection, and fidelity constraints. & Content Fidelity, Instruction Preservation, and failure analysis. \\
  \midrule
  Specialists require reviewability & Generated errors may be hard for learners to detect. & Use Evaluator--Refiner verification and expose failed constraints as risk flags. & Ablation, refinement-budget sensitivity, and human-review boundary. \\
  \midrule
  Materials need domain validity & Arbitrary questions may not measure the intended construct. & Use expert-curated materials and avoid rewriting when language form is the tested construct. & Expert curation counts and non-use cases. \\
  \midrule
  Over-cueing can become visual overload & Highlighting every important word may distract readers or look like answer hinting. & Use a conservative highlight budget, emphasize labels rather than full question text, and prefer generous spacing. & Expert spot-check comments and follow-up audit. \\
  \midrule
  High-stakes outputs need accountability & A fluent rewrite can still hide a harmful change. & Preserve a trace from stakeholder requirement to blueprint, protected spans, verification result, and human-review flag. & Traceability table, Evaluator--Refiner diagnostics, and deployment boundary. \\
  \bottomrule
  \end{tabular}
  \end{table*}

  \subsection{Failure Modes and Non-Use Cases}

  We use failure analysis as a complement to aggregate DCFI scores. The most important failure classes are: (i) \emph{over-simplification}, where a model removes a domain term that remains necessary for the task; (ii) \emph{instruction drift}, where a question asks for a different operation after rewriting; (iii) \emph{hallucinated scaffolding}, where the model adds an explanation or hint not present in the original; (iv) \emph{formatting-only improvement}, where spacing improves readability but linguistic burden remains; and (v) \emph{unnecessary rewriting}, where the text form itself is part of what the task measures. These cases motivate the risk flag: outputs that repeatedly fail fidelity or instruction checks should be reviewed rather than deployed automatically.

  DFT-GEN should not be used as a diagnosis or treatment tool, and should not automatically rewrite texts when wording, style, ambiguity, or difficulty is itself the construct being assessed. Examples include language-proficiency reading tests, literary interpretation, poetry, grammar questions, and rhetorical-analysis tasks. In those settings, we recommend layout-only accommodations such as spacing, segmentation, and optional highlighting, because they reduce visual burden without changing the linguistic construct.

  \subsection{Same-Backbone Strong Baseline Definitions}
  \label{app:strong_baselines}

  To avoid comparing DFT-GEN only against weak single-pass prompting, we define four same-backbone controls that use the same underlying LLM as the corresponding DFT-GEN run. \textbf{Self-refine} follows iterative self-feedback ideas~\cite{madaan2023selfrefine}: the model produces an initial dyslexia-friendly rewrite, critiques its own output for meaning loss and readability issues, and revises once. \textbf{Critic-refine} follows critique-and-revision agent patterns related to reflection-based LLM control~\cite{shinn2023reflexion}: separate critic and rewriter calls are used, but without the sentence-level Architect blueprint or placeholder map. \textbf{Structured constraint prompt} gives the model a checklist of named entities, numbers, task verbs, and source references, but asks for a single final rewrite. \textbf{Multi-pass prompt} performs sequential passes for segmentation, wording simplification, and formatting, but does not include the Evaluator--Refiner PASS/FAIL gate. For the planned five-excerpt pilot extension, we additionally prepare an ACCESS-inspired controllable simplification baseline~\cite{martin2020access} that controls sentence length, chunking, lexical complexity, and structure without DFT-GEN's traceable controller. These controls test whether DFT-GEN's gains come from the specific traceable architecture rather than merely from spending more LLM calls or writing a longer prompt.

  \subsection{Targeted Expert Audit Protocol}
  \label{app:expert_audit}

  Our expert evidence has two layers. The first layer consists of five formative expert interviews or written-feedback exchanges with dyslexia/SEN/accessibility practitioners; these shaped the requirements, deployment boundaries, and the DAC rules. The second layer is a targeted English output audit, which is narrower and is used for output-quality calibration rather than for claiming broad expert consensus. Since the available output rater can read English outputs, this audit is intentionally English-only and does not claim to validate the Chinese setting. Across the two rounds, the audit contains 80 samples and 160 candidate-output ratings. The first round used 20 English items in blind A/B form. After decoding the A/B key, DFT-GEN was preferred in 14 of 20 comparisons; it received Accept on 8 items and Accept or Revise on 18 of 20 items. The expert's comments rewarded clear source/context separation, short visual information units, explicit question blocks, preserved references and task verbs, and generous spacing. The main criticism was visual overload from too much highlighting, especially when cues looked like answer hints.

  We therefore conducted a follow-up expert audit with 60 additional samples and 120 candidate-output ratings. Thirty samples come from existing English materials; thirty are synthetic exam-style stress-test items spanning history, biology, and geography. The synthetic items are used only to stress-test formatting and fidelity risks under controlled question patterns; they are not reported as authentic exam data. Each sample shows the original text and two anonymized candidate outputs in randomized A/B order. The form asks for 1--5 meaning preservation, 1--5 dyslexia-friendly presentation, Accept/Revise/Reject deployment decisions for each candidate, an A/B/Tie preference, and optional comments. In the follow-up round, the revised DFT-GEN candidate was preferred in 52 of 60 comparisons and accepted as-is in 50 of 60 cases, with mean scores of 4.52/5 for meaning preservation and 4.18/5 for dyslexia-friendly presentation. Remaining comments were mostly local presentation edits rather than systematic meaning loss.

  \begin{table*}[t]
  \centering
  \small
  \caption{Targeted 80-sample expert audit protocol. The follow-up audit extends coverage for qualitative failure analysis rather than replacing controlled user evaluation.}
  \label{tab:expert_audit_protocol}
  \begin{tabular}{p{3.0cm}p{4.2cm}p{7.0cm}}
  \toprule
  \textbf{Audit stage} & \textbf{Material} & \textbf{Purpose and ratings} \\
  \midrule
  First-round blind spot-check & 20 existing English items; 40 candidate-output ratings & Compare anonymized candidate rewrites; record meaning preservation, dyslexia-friendly presentation, deployment decision, preference, and comments. Used to identify expert-valued patterns and failure modes. \\
  \midrule
  Follow-up expert audit & 30 existing English items + 30 synthetic stress-test items; 120 candidate-output ratings & Test whether lighter highlighting, generous spacing, and explicit question blocks address prior expert concerns. Synthetic items are transparently marked as stress tests and are not treated as real exam samples. \\
  \bottomrule
  \end{tabular}
  \end{table*}

  \begin{table}[t]
  \centering
  \small
  \caption{Overall English expert-audit results.}
  \label{tab:expert_rating_summary}
  \begin{tabular}{lc}
  \toprule
  \textbf{Metric} & \textbf{Result} \\
  \midrule
  Meaning preservation & 4.32/5 \\
  \midrule
  Dyslexia-friendly presentation & 3.94/5 \\
  \midrule
  Accepted as-is & 62/80 \\
  \midrule
  Accept or revise & 74/80 \\
  \midrule
  Preferred overall & 66/80 \\
  \bottomrule
  \end{tabular}
  \end{table}

  This audit is not a substitute for a large controlled study, but it provides a practical safety check focused on the expert concerns that motivated the framework: lost terminology, added hints, changed instructions, over-highlighting, and visually crowded formatting.

  \paragraph{Expert-score correlation analysis.}
The 80-sample audit export provides meaning preservation (1--5), dyslexia-friendly presentation (1--5), deployment decision (Accept/Revise/Reject), A/B/Tie preference, and comments for each candidate output. We map Accept/Revise/Reject to an ordinal deployment score and compute Pearson and Spearman correlations between candidate-level diagnostics and expert signals. Table~\ref{tab:expert_dcfi_correlation_appendix} shows the intended division of labor. DCFI has only a weak positive association with expert meaning scores, which is expected for a thresholded fidelity and utility screen over mostly acceptable candidates; it is meant to flag risky outputs, not replace expert semantic grading. B-DVAS-VL is much more aligned with expert presentation and A/B preference, supporting its use as a rendered visual-accessibility diagnostic.

  \begin{table*}[t]
  \centering
  \small
  \setlength{\tabcolsep}{3pt}
  \caption{Candidate-level correlations between automatic diagnostics and expert ratings over the 80-sample blind audit ($160$ A/B candidate ratings).}
  \label{tab:expert_dcfi_correlation_appendix}
  \begin{tabular}{llcc}
  \toprule
  \textbf{Predictor} & \textbf{Expert signal} & \textbf{Pearson $r$} & \textbf{Spearman $\rho$} \\
  \midrule
  DCFI & Meaning preservation & 0.222 & 0.220 \\
  \midrule
  B-DVAS-VL & Dyslexia-friendly presentation & 0.650 & 0.705 \\
  \midrule
  B-DVAS-VL & A/B preference & 0.475 & 0.518 \\
  \midrule
  Content fidelity & Meaning preservation & 0.256 & 0.174 \\
  \midrule
  Effectiveness & Meaning preservation & 0.075 & 0.124 \\
  \midrule
  Punctual/structural fidelity & Meaning preservation & 0.139 & 0.130 \\
  \bottomrule
  \end{tabular}
  \end{table*}

  \subsection{Reliability Controls for Human and Expert Evaluation}
  \label{app:reliability_controls}

  We treat the current human and expert evidence as a controlled pilot plus targeted expert calibration, not as a definitive clinical or educational trial. A proportionate extension is a five-excerpt set spanning history, biology, and geography. This expansion is not meant to assess curriculum knowledge; the measured construct remains information access under reading burden. Excerpts are selected before model output inspection and screened for comparable task type, approximate length band, information density, and presence of explicit answerable evidence. The goal is not to make all texts identical, but to avoid a situation where the observed speed or effort gains are driven by two unusually easy passages or by evidence that happens to be located in an obvious position.

  \begin{table*}[t]
  \centering
  \small
  \setlength{\tabcolsep}{3pt}
  \caption{Planned five-excerpt extension for a broader information-access pilot. Ex1--Ex2 are the current pilot excerpts; Ex3--Ex5 add typical source structures rather than harder subject knowledge.}
  \label{tab:five_excerpt_extension}
  \begin{tabular}{p{1.0cm}p{2.0cm}p{4.1cm}p{6.1cm}}
  \toprule
  \textbf{Ex.} & \textbf{Domain} & \textbf{Source structure} & \textbf{Reason for inclusion} \\
  \midrule
  Ex1 & History & Two-source prompt with several named concepts and task anchors. & Tests whether DFT-GEN preserves source references and required operations while reducing visual load. \\
  \midrule
  Ex2 & History & Single-source causal/explanatory prompt with numerical details. & Tests evidence retrieval when key facts are embedded in a dense continuous paragraph. \\
  \midrule
  Ex3 & Biology & Concept/mechanism explanation with terminology and process steps. & Tests whether accessibility rewriting preserves domain terms while making sequential information easier to track. \\
  \midrule
  Ex4 & Geography & Data- or map-style description converted to textual evidence. & Tests access to numbers, comparisons, and spatial or tabular cues without turning them into answer hints. \\
  \midrule
  Ex5 & Mixed social-science / public-information text & Multi-paragraph passage with evidence distributed across sections. & Tests whether readers can locate information when the answer is not adjacent to the question cue. \\
  \bottomrule
  \end{tabular}
  \end{table*}

  For each excerpt, candidate presentations are generated under the same content constraints and then assigned condition labels that do not reveal system identity. Participants and output raters see generic condition names rather than model names such as DFT-GEN, strong-prompt, or commercial tool. For A/B expert review, candidate order is randomized per item and the A/B key is stored separately until after ratings are exported. For multi-condition user tasks, excerpt order and condition order are counterbalanced as far as the sample size allows; when full counterbalancing is not possible, assignment is blocked so that no condition always appears first or last.

  We also control visual presentation separately from linguistic content. All conditions are rendered in the same interface with the same font family, font size, line spacing, page width, background, and interaction affordances unless the condition being tested explicitly concerns layout. This prevents a model from appearing better because it received a more favorable UI. Learning effects are reduced by using different but comparable excerpts across conditions, limiting repeated exposure to the same factual content, and recording order so that analyses can check whether later trials are faster because of practice rather than text quality.

  \begin{table*}[t]
  \centering
  \small
  \setlength{\tabcolsep}{3pt}
  \caption{Evaluation controls added to address sample-size and reliability concerns. The full materials, anonymized expert summaries, and written-feedback extracts should be placed in the appendix or artifact package where privacy permits.}
  \label{tab:evaluation_controls_appendix}
  \begin{tabular}{p{3.1cm}p{5.2cm}p{5.2cm}}
  \toprule
  \textbf{Threat to reliability} & \textbf{Control in protocol} & \textbf{Appendix / artifact material} \\
  \midrule
  Too few excerpts & Expand from two pilot excerpts to five typical excerpts covering source structure, length, domain terminology, numerical evidence, and evidence location. & Excerpt inventory with subject, length band, question type, and reason for inclusion. \\
  \midrule
  Expert-number ambiguity & Separate five formative expert interviews / written-feedback exchanges from the narrower output spot-check. & Anonymized expert matrix: role, experience band, feedback mode, topics covered, and design implication. \\
  \midrule
  Condition identity bias & Use generic condition labels and blind A/B ordering for expert output review. & A/B randomization seed, exported key file, and rating form screenshots or text specification. \\
  \midrule
  Text length and information imbalance & Pre-screen excerpts by approximate length, number of source references, marks, task verbs, and evidence density. & Table reporting token/word length, number of questions, marks, source references, and subject. \\
  \midrule
  Visual presentation confound & Use the same interface, font, size, spacing, width, and background for all non-layout conditions. & Interface specification and CSS/settings summary. \\
  \midrule
  Reading-order and learning effects & Counterbalance or block-randomize excerpt and condition order; log presentation order for analysis. & Assignment table and order log template. \\
  \midrule
  Output quality randomness & Use fixed model versions, temperature, prompt templates, and decoding settings; cache outputs before evaluation. & Model/version table, prompt files, and cached output hashes or identifiers. \\
  \midrule
  Unverifiable expert feedback & Store anonymized written-feedback extracts and interview summaries separately from raw identifiable records. & Expert-feedback appendix with de-identified quotations, comments, and mapping to DAC rules. \\
  \bottomrule
  \end{tabular}
  \end{table*}

  \begin{table*}[t]
  \centering
  \scriptsize
  \setlength{\tabcolsep}{3pt}
  \caption{Anonymized expert and written-feedback extracts used to justify design and evaluation controls. Wording is lightly normalized; identifying names, institutions, and locations are withheld for review.}
  \label{tab:expert_feedback_extracts}
  \begin{tabular}{p{1.1cm}p{3.0cm}p{5.2cm}p{4.8cm}}
  \toprule
  \textbf{ID} & \textbf{Source type} & \textbf{Feedback extract or summary} & \textbf{Resulting design / evaluation control} \\
  \midrule
  E1 & Dyslexia / reading specialist & Dyslexic readers may understand the underlying concept but need more time to decode, sequence, and track dense written information. & Treat reading burden as processing friction; measure time and effort, not only correctness. \\
  \midrule
  E2 & SEN / school-support practitioner & Shorter text is not always better if it removes wording that learners need for the task or later classroom discussion. & Preserve domain terms and task verbs; report content and instructional fidelity. \\
  \midrule
  E3 & Learning-support leadership & AI-generated accommodations should be reviewable, especially in high-stakes educational contexts where small wording changes can alter the task. & Add Evaluator--Refiner diagnostics, risk flags, and a human-review boundary. \\
  \midrule
  E4 & Accessibility-oriented educator & Crowded presentation, weak spacing, and over-cueing can increase fatigue even when the language is simpler. & Add DAC visual-load calibration, generous spacing, and highlight-budget controls. \\
  \midrule
  E5 & Blind output rater & Strong outputs separated source/context from questions, used short visual units, kept marks and source references visible, and avoided highlighting whole question strings. & Derive DAC question--task separation, source-anchor preservation, and follow-up ablation diagnostics. \\
  \midrule
  E5 & Blind output rater & Poor outputs either diluted subject terminology or became visually overwhelming through excessive emphasis. & Interpret pilot and expert results jointly: answerability requires fidelity, while presentation quality requires separate expert/DAC checks. \\
  \bottomrule
  \end{tabular}
  \end{table*}

  \begin{table*}[t]
  \centering
  \scriptsize
  \setlength{\tabcolsep}{3pt}
  \caption{Anonymized summaries from four formative adult user interviews. These summaries report coded themes rather than identifiable biographies; one additional user profile was not yet transcribed and is therefore not included here.}
  \label{tab:user_interview_summaries}
  \begin{tabular}{p{1.0cm}p{3.0cm}p{4.4cm}p{3.4cm}p{3.2cm}}
  \toprule
  \textbf{ID} & \textbf{Reading profile} & \textbf{Reported barriers} & \textbf{Helpful support described} & \textbf{Design implication} \\
  \midrule
  U1 & Adult with formally diagnosed or long-standing dyslexia-related reading difficulty. & Dense paragraphs made it hard to keep place; long sentences increased rereading and fatigue even when the topic itself was understandable. & Clear paragraph breaks, shorter visual units, and the ability to scan one idea at a time. & Segment dense text without assuming lack of conceptual ability. \\
  \midrule
  U2 & Adult who self-identified persistent reading difficulties and passed the screening/plausibility checks. & Unfamiliar terms were not always the main problem; the harder part was tracking which details belonged to the question or task. & Explicit labels, visible source references, and task wording separated from background information. & Preserve task structure and make question blocks explicit. \\
  \midrule
  U3 & Adult reader reporting slow reading speed and high effort under time pressure. & Time pressure amplified decoding difficulty; crowded text and low visual contrast increased stress and avoidance. & Generous spacing, stable layout, and reduced visual clutter rather than aggressive rewriting. & Add visual-load calibration and avoid over-cueing. \\
  \midrule
  U4 & Adult reader who valued access to authentic terminology and did not want simplified text to feel infantilizing. & Over-simplified summaries could feel easier but risked losing useful subject vocabulary and confidence in the original material. & Keep important terms visible while simplifying surrounding structure and sentence flow. & Preserve domain terms and use fidelity checks before deployment. \\
  \bottomrule
  \end{tabular}
  \end{table*}

  \subsection{Pilot Stimuli and Administration Details}
  \label{app:pilot_details}

  This appendix records the concrete pilot materials and administration choices so that the near-ceiling DFT-GEN correctness result can be interpreted appropriately. The pilot was designed as a controlled information-access and reading-burden check, not as a comprehensive history achievement test. The five comprehension questions for each excerpt were intentionally simple evidence-retrieval items whose answers appeared in the passage. They were identical across all six conditions for the same excerpt, and the researcher used the same oral script, timing rule, rating questions, and recording template across conditions.

  \begin{table*}[t]
  \centering
  \small
  \setlength{\tabcolsep}{3pt}
  \caption{Pilot excerpt inventory and condition arms. The original file names are anonymized to preserve review anonymity while retaining enough information to audit the task design.}
  \label{tab:pilot_excerpt_inventory}
  \begin{tabular}{p{2.2cm}p{4.5cm}p{4.8cm}p{3.5cm}}
  \toprule
  \textbf{Excerpt} & \textbf{Anonymized source ID} & \textbf{Topic and source structure} & \textbf{Condition arms} \\
  \midrule
  Ex1 & 2024 national history exam item, blank paper, item 1 & Two-source prompt on the development of Chinese mathematics from antiquity to the modern period; three original open-ended exam questions. & Original, DFT-GEN, ATS, StrongPrompt, LayoutOnly, DyslexiaHelper. \\
  \midrule
  Ex2 & 2024 provincial history exam item, blank paper, item 2 & Single-source prompt on postwar international monetary-system change, the dollar, the euro, and quantitative easing; two original open-ended exam questions. & Original, DFT-GEN, ATS, StrongPrompt, LayoutOnly, DyslexiaHelper. \\
  \bottomrule
  \end{tabular}
  \end{table*}

  \begin{table*}[t]
  \centering
  \small
  \setlength{\tabcolsep}{3pt}
  \caption{Pilot condition definitions and expected risks. All condition labels shown to participants were generic and did not reveal system identity.}
  \label{tab:pilot_condition_definitions}
  \begin{tabular}{p{2.4cm}p{6.0cm}p{5.8cm}}
  \toprule
  \textbf{Condition} & \textbf{Presentation shown to participants} & \textbf{Reason for inclusion} \\
  \midrule
  Original & Source text and original task wording without accessibility rewriting. & Baseline for authentic exam-style reading burden and answerability. \\
  \midrule
  DFT-GEN & Constraint-preserving rewrite generated by the DFT-GEN pipeline with short visual units and preserved task requirements. & Tests whether reduced burden can be achieved without deleting answer-critical evidence. \\
  \midrule
  ATS & Neural automatic text simplification output from a local baseline. & Tests whether generic simplification can become fast but lossy or malformed. \\
  \midrule
  StrongPrompt & Same-backbone prompted rewrite with sentence splitting and plain-language instructions but without full DFT-GEN verification. & Controls for whether a strong prompt alone explains the gains. \\
  \midrule
  LayoutOnly & Original wording retained with spacing and segmentation only. & Controls for visual layout benefits without linguistic rewriting. \\
  \midrule
  DyslexiaHelper & Short summary-style helper output produced by an existing dyslexia-oriented assistance style. & Tests a high-compression accommodation that may feel easy but remove task-critical information. \\
  \bottomrule
  \end{tabular}
  \end{table*}

  \begin{table*}[t]
  \centering
  \scriptsize
  \setlength{\tabcolsep}{3pt}
  \caption{Exact comprehension checks and answer keys used in the pilot. These were evidence-retrieval checks rather than open-ended essay grading rubrics.}
  \label{tab:pilot_comprehension_questions}
  \begin{tabular}{p{1.0cm}p{6.5cm}p{6.6cm}}
  \toprule
  \textbf{Ex.} & \textbf{Question shown after reading} & \textbf{Accepted answer} \\
  \midrule
  Ex1 & Q1: The Han-dynasty mathematical book mentioned in Material 1 is \_\_\_\_. & \textit{Nine Chapters on the Mathematical Art} (Jiu Zhang Suan Shu). \\
  \midrule
  Ex1 & Q2: Name any two world-class mathematicians mentioned in Material 1. & Any two of Liu Hui, Zu Chongzhi, and Zu Geng. \\
  \midrule
  Ex1 & Q3: What institution was established in the Sui--Tang period to teach mathematical classics? & Mathematical Studies (Suanxue). \\
  \midrule
  Ex1 & Q4: During the Self-Strengthening Movement, what did the Qing court add to the Tongwen Guan? & The Astronomy and Mathematics Division. \\
  \midrule
  Ex1 & Q5: In 1940, the Chinese Communist Party established the Academy of Natural Sciences in \_\_\_\_. & Yan'an. \\
  \midrule
  Ex2 & Q1: The passage discusses the period from the 1970s to before the birth of the euro. Fill in the decade. & 1970s / 70. \\
  \midrule
  Ex2 & Q2: Which factor is described as having a one-way impact on European currencies? & Dollar exchange-rate fluctuation. \\
  \midrule
  Ex2 & Q3: What were the dollar's global foreign-reserve shares in 1999 and 2008? & 72.7\% and 64\%. \\
  \midrule
  Ex2 & Q4: What state did the international monetary system enter? & Dollar--euro dual-dominant currency system. \\
  \midrule
  Ex2 & Q5: Under dual dominance, what happens if one dominant currency is over-issued? & The other dominant currency follows, and the effect spreads to small and medium currencies. \\
  \bottomrule
  \end{tabular}
  \end{table*}

  \begin{table*}[t]
  \centering
  \small
  \setlength{\tabcolsep}{3pt}
  \caption{Pilot administration script and timing rule. The same script was used across all conditions.}
  \label{tab:pilot_script}
  \begin{tabular}{p{3.4cm}p{10.2cm}}
  \toprule
  \textbf{Step} & \textbf{Script or logging rule} \\
  \midrule
  Opening instruction & ``I will show you a reading passage. Please read at your own pace. After you finish, I will ask five simple comprehension questions. You may ask me to return to an earlier part of the passage, but please try to read it independently first.'' \\
  \midrule
  Timing start & The researcher says: ``I am starting the timer now; you may begin reading.'' Timing starts when the passage is visible and the instruction is complete. \\
  \midrule
  Timing stop & Timing stops after the participant answers the fifth comprehension question. If the participant asks to revisit the passage, the time continues running. \\
  \midrule
  Post-task ratings & The researcher asks three 1--7 oral rating questions: reading effort (1 = no effort, 7 = very effortful), willingness to use this version in future tasks (1 = not willing, 7 = very willing), and answer confidence (1 = not confident, 7 = very confident). \\
  \midrule
  Optional comment & The researcher asks one short open question: ``What felt most difficult: unfamiliar words, long sentences, or dense information?'' Notes are recorded only in anonymized form. \\
  \bottomrule
  \end{tabular}
  \end{table*}

  \begin{table*}[t]
  \centering
  \scriptsize
  \setlength{\tabcolsep}{3pt}
  \caption{Pilot data-recording template. Participant identifiers are pseudonymous and stored separately from any contact information.}
  \label{tab:pilot_record_template}
  \resizebox{\textwidth}{!}{%
  \begin{tabular}{p{2.0cm}p{1.6cm}p{1.9cm}p{1.5cm}p{1.5cm}p{1.4cm}p{2.8cm}p{1.4cm}p{1.6cm}p{1.5cm}p{2.0cm}}
  \toprule
  \textbf{Participant} & \textbf{Excerpt} & \textbf{Condition} & \textbf{Start} & \textbf{End} & \textbf{Seconds} & \textbf{Q1--Q5 responses} & \textbf{Correct/5} & \textbf{Effort} & \textbf{Willing.} & \textbf{Notes} \\
  \midrule
  Pseudonymous ID & Ex1/Ex2 & Generic arm label & Time stamp & Time stamp & Total time & Raw short answers or option letters & 0--5 & 1--7 & 1--7 & Optional anonymized comment \\
  \bottomrule
  \end{tabular}
  }
  \end{table*}

  The near-ceiling DFT-GEN correctness should therefore be read narrowly. It does not show that participants achieved perfect historical reasoning or that the excerpts were sufficient for population-level validation. It shows that, for these two evidence-retrieval tasks, DFT-GEN preserved the information needed to answer the questions while reducing time and perceived effort. The more general claim requires the planned five-excerpt extension with more varied source structures, evidence locations, and order counterbalancing, plus additional expert review.

  \begin{table*}[t]
  \centering
  \small
  \setlength{\tabcolsep}{3pt}
  \caption{Anonymized expert evidence matrix for the five formative expert sources. Exact names, institutions, and raw transcripts are withheld during review; role summaries and design implications are reported for auditability.}
  \label{tab:five_expert_matrix}
  \begin{tabular}{p{1.3cm}p{3.1cm}p{2.2cm}p{4.0cm}p{3.6cm}}
  \toprule
  \textbf{ID} & \textbf{Role / expertise} & \textbf{Feedback mode} & \textbf{Topics covered} & \textbf{Influence on DFT-GEN} \\
  \midrule
  E1 & Dyslexia / reading specialist & Interview / written notes & Processing time, authentic wording, classroom accommodations. & Preserve terminology; avoid infantilizing simplification. \\
  \midrule
  E2 & SEN teacher / school-support practitioner & Interview & Undiagnosed learners, classroom workload, assistive technology use. & Make output easy to scan and practical for teacher review. \\
  \midrule
  E3 & Learning-support leadership / dyslexia association experience & Interview & Under-identification, stigma, human review, deployment risk. & Add risk flags and conservative deployment boundary. \\
  \midrule
  E4 & Accessibility-oriented educator / practitioner & Written feedback & Visual stress, spacing, cue overload, readable presentation. & Add DAC visual-load calibration and highlight budget. \\
  \midrule
  E5 & Dyslexia therapist / accessibility practitioner & Interview / written feedback & Learner confidence, safe AI support, reviewability, and preserving task anchors. & Add traceability, source/task anchors, and review gates. \\
  \bottomrule
  \end{tabular}
  \end{table*}

  \begin{table}[H]
  \centering
  \small
  \setlength{\tabcolsep}{3pt}
  \caption{Impact of maximum refinement budget $L$ on a sensitivity subset.}
  \label{tab:l_sensitivity_appendix}
  \begin{tabular}{lccccc}
  \toprule
  \textbf{$L$} & \textbf{DCFI} & \textbf{Acc.} & \textbf{Task} & \textbf{Safe} & \textbf{Time (s)} \\
  \midrule
  1 & \textbf{0.499} & 0.876 & 0.834 & \textbf{1.000} & \textbf{22.4} \\
  2 & 0.414 & \textbf{0.934} & \textbf{0.852} & \textbf{1.000} & 29.0 \\
  3 & 0.429 & 0.913 & 0.813 & \textbf{1.000} & 35.2 \\
  \bottomrule
  \end{tabular}
  \end{table}

  The DCFI ranking is stable under stakeholder-informed weight alternatives (Spearman $\rho \geq 0.976$), suggesting that the main conclusions do not depend on a single exact weight setting.

  \begin{table}[H]
  \centering
  \footnotesize
  \setlength{\tabcolsep}{3pt}
  \caption{Impact of DCFI component weights. Corr. denotes Spearman's $\rho$ with the default ranking.}
  \label{tab:weight_sensitivity_appendix}
  \resizebox{\columnwidth}{!}{%
  \begin{tabular}{lcc}
  \toprule
  \textbf{Configuration} & \textbf{Weights $(w_a,w_c,w_t,w_s)$} & \textbf{Corr.} \\
  \midrule
  Default (fidelity-dominant) & (0.20, 0.35, 0.30, 0.15) & 1.000 \\
  Accessibility-heavy & (0.40, 0.25, 0.20, 0.15) & 0.976 \\
  Balanced & (0.30, 0.30, 0.25, 0.15) & 0.988 \\
  Task-heavy & (0.15, 0.30, 0.40, 0.15) & 0.988 \\
  Uniform & (0.25, 0.25, 0.25, 0.25) & 1.000 \\
  \bottomrule
  \end{tabular}
  }
  \end{table}

  \begin{table}[H]
  \centering
  \scriptsize
  \setlength{\tabcolsep}{2pt}
  \caption{Directional alignment between DCFI signals and human A/B outcomes on two excerpts. Here $A$ is Original and $B$ is DFT-GEN.}
  \label{tab:dcfi_human_alignment_appendix}
  \resizebox{\columnwidth}{!}{%
  \begin{tabular}{lcccccccccc}
  \toprule
  \textbf{Ex} & \textbf{DCFI$_A$} & \textbf{DCFI$_B$} & \textbf{Eff$_A$} & \textbf{Eff$_B$} & \textbf{Time$_A$} & \textbf{Time$_B$} & \textbf{Acc$_A$} & \textbf{Acc$_B$} & \textbf{Effort$_A$} & \textbf{Effort$_B$} \\
  \midrule
  Ex1 & 0.692 & 0.911 & 0.000 & 0.850 & 290.2 & 207.0 & 3.4 & 4.7 & 6.1 & 2.2 \\
  Ex2 & 0.700 & 0.943 & 0.000 & 0.850 & 299.5 & 198.3 & 3.4 & 4.8 & 6.1 & 2.0 \\
  \bottomrule
  \end{tabular}
  }
  \end{table}

  \subsection{Reproducibility, Assets, and Human-Study Details}

  The code and artifact package is available at \url{https://github.com/MorrisYUJQ/DFT-GEN}. The experiments do not train a new model. They call existing API-accessible LLMs, including DeepSeek-V3, Gemini-2.5-Pro, Qwen-3-Max, GPT-4o, and Claude-4-Sonnet, and run local evaluation scripts for formatting checks, DCFI aggregation, ablations, and statistical summaries. Each DFT-GEN item normally requires an Architect call, a Writer call, and an Evaluator call, with up to two additional refinement passes for flagged outputs. For the 2,280-item automatic benchmark and five LLM backbones, this corresponds to roughly $3.4\times 10^4$--$5.7\times 10^4$ generation/verification API calls, depending on the number of refinements, plus lightweight local scoring. No GPU training is required; the local scripts can be run on a standard CPU workstation, while wall-clock time is dominated by API latency and rate limits.

  The artifact package is intended to contain: prompt templates, model/version settings, example inputs and outputs, DCFI scoring code, baseline scripts, ablation and sensitivity scripts, and README instructions for regenerating the tables. Raw interview transcripts and identifiable participant-level records are excluded from the public package; de-identified summaries and item-format examples are sufficient for auditing the method and reproducing aggregate analyses.

  Existing models and APIs are accessed through their official terms of use, and the corresponding model reports or system cards are cited in the paper. Public benchmark or source materials are credited where used. Newly collected interviews, expert-curation records, and participant-level pilot materials are treated as research artifacts: we release only anonymized summaries and de-identified examples, and plan to release broader derived materials subject to privacy and copyright constraints.

  For the pilot study, participants were informed that the task was a non-clinical reading-usability study rather than diagnosis, treatment, or educational assessment. They read passages under assigned presentation conditions, answered comprehension questions, and reported perceived effort and willingness-to-use ratings. Presentation conditions were labeled generically rather than by system name, and excerpt order was controlled to reduce ordering effects. The comprehension questions were intentionally tied to explicit evidence in the passage; this makes the task suitable for testing whether rewriting preserves answerability, but it also makes ceiling effects possible when the relevant evidence is made easy to locate. The task interface consisted of text passages, comprehension items, and rating questions, so screenshots are not necessary to understand the procedure; the released artifact package provides the task structure and item format in text form.

  \section*{Generative AI Usage}
  This paper uses third-party LLM APIs as system components for generation, verification, and an LLM-based proxy signal in DCFI. All claims, writing, analysis, and anonymization remain the authors' responsibility.

  \end{document}


\maketitle

\section{What this file is}
This PDF is intended to be packaged as the submission supplementary ZIP (separate from the main paper PDF). It contains extended methodological details, full tables/figures, and additional analyses that do not fit in the 7+2 page limit.

\section{Formative Study: Recruitment, Screening, and Protocol}
\paragraph{Recruitment.}
We recruited 20 adults via social media who self-identified as having dyslexia or persistent reading difficulties (including some with prior formal diagnosis).

\paragraph{Pre-screening and response-time sanity check.}
To reduce false positives, we used an adult dyslexia screening checklist (Smythe \& Siegel, 2004) as an initial filter. We additionally used response-time logs from an online reading task as a conservative plausibility check. The total text volume in the task was approximately $508$ Chinese characters (including prompts and short instructions). We set an upper-bound reading rate of $10$ characters/second ($600$ chars/min) and approximated interaction overhead as $19$ clicks $\times$ $1$ second each, yielding a minimum plausible completion time of $508/10 + 19 \approx 70$ seconds. Respondents substantially faster than this threshold were excluded.

\paragraph{Educator consultation.}
We additionally consulted an educator based in Macao with practical experience supporting dyslexic learners. Semi-structured interviews were approximately 30 minutes, audio-recorded with consent, transcribed, and thematically summarized. Raw transcripts are not publicly released due to anonymization commitments; de-identified excerpts or aggregated coding summaries can be made available upon reasonable request.

\subsection{Interview participant demographics (anonymized)}
Table~\ref{tab:interview_demographics} summarizes the demographics and self-reported dyslexia-related reading characteristics of the 20 interview participants. To avoid domain-specific leakage, we remove context-specific barriers and retain only general reading-related issues (e.g., visual density, jargon, low contrast, and information overload).

\begin{table*}[t]
\centering
\scriptsize
\setlength{\tabcolsep}{2pt}
\renewcommand{\arraystretch}{1.15}
\caption{Interview participant demographics and self-reported reading-related characteristics (anonymized). Barrier descriptions are lightly normalized to remove domain-specific context while preserving reading-related issues.}
\label{tab:interview_demographics}
\resizebox{\textwidth}{!}{%
\begin{tabular}{lllllll p{6.2cm} p{7.6cm}}
\toprule
\textbf{ID} & \textbf{Gender} & \textbf{Age} & \textbf{Education} & \textbf{Occupation} & \textbf{Diagnosis} & \textbf{Status} & \textbf{Dyslexia-related characteristics (self-reported)} & \textbf{Reading barriers (generalized)} \\
\midrule
P01 & Male & 31 & Bachelor & Community worker & Self-identified & -- & Skipping; misinterpretation; repetition; slowness; effortfulness & Information overload; jargon/terminology; visual density \\
P02 & Male & 25 & Bachelor & Product manager & Self-identified & -- & Slowness; distraction; repetition; word-by-word decoding; sequential memory difficulty & Information overload; small fonts; visual density; technical jargon \\
P03 & Female & 20 & Bachelor & Student & Self-identified & -- & Transposition; slowness; skipping; misreading; selective avoidance & Language barrier; jargon/terminology; visual density; sequence confusion \\
P04 & Female & 23 & Master & Freelancer & Self-identified & -- & Skipping; misinterpretation; repetition; inattentiveness; effortfulness & Jargon/terminology; information overload; technical terms \\
P05 & Male & 26 & Bachelor & Real estate planner & Self-identified & -- & Distraction; repetition; vocalization; effortfulness & Text density; jargon/terminology \\
P06 & Female & 21 & Bachelor & Student & Formally diagnosed & -- & Slowness; effortfulness; incomprehension; visual distortion; fatigue & Layout clutter; small fonts; information overload; jargon/terminology \\
P07 & Male & 30 & Bachelor & Entrepreneur & Self-identified & -- & Effortfulness; slowness; comprehension difficulty & Jargon/terminology; comprehension difficulty; social pressure \\
P08 & Female & 26 & Master & Freelancer & Self-identified & -- & Skipping; inattentiveness; effortfulness; misreading & Information overload; jargon/terminology; visual density; obscure vocabulary \\
P09 & Female & 25 & Bachelor & Project manager & Self-identified & -- & Skipping; repetition; inattentiveness; effortfulness & Visual density; small fonts; low contrast; text-heaviness; psychological resistance \\
P10 & Female & 20 & Bachelor & Student & Formally diagnosed & -- & Slowness; misinterpretation; fatigue; incomprehension; effortfulness & Visual density; text density; jargon/terminology; social pressure \\
P11 & Female & 29 & Master & Self-employed & Formally diagnosed & -- & Slowness; misinterpretation; fatigue; forgetting; word-by-word reading & Visual density; small fonts; information overload; jargon/terminology \\
P12 & Female & 25 & Doctoral & Student & Self-identified & -- & Skipping; repetition; inattentiveness; visual distortion; effortfulness & Visual density; small fonts; information overload; jargon/terminology \\
P13 & Male & 26 & Master & Township staff & Self-identified & -- & Slowness; repetition; distraction; fatigue; low retention; complexity sensitivity & Information overload; visual density; comprehension difficulty; jargon/terminology \\
P14 & Female & 23 & Master & Student & Self-identified & -- & Skipping; slowness; repetition; inattentiveness; effortfulness & Information overload; comprehension difficulty; jargon/terminology \\
P15 & Female & 28 & Master & Teacher & Self-identified & -- & Slowness; skipping; fatigue; incomprehension; effortfulness & Visual density; information overload; jargon/terminology; difficulty seeking help \\
P16 & Male & 30 & Master & Engineer & Formally diagnosed & -- & Effortfulness; incomprehension; inattentiveness; fragmented reading; fatigue & Visual density; information overload; multilingual formatting; focus difficulty \\
P17 & Male & 22 & Bachelor & Teacher & Self-identified & -- & Slowness; distraction; fatigue; jargon difficulty; conceptual struggle & Small fonts; visual density; jargon/terminology \\
P18 & Female & 27 & Bachelor & Manufacturing staff & Self-identified & -- & Slowness; distraction; incomprehension; fatigue & Visual density; information overload; small fonts; jargon/terminology \\
P19 & Male & 24 & Master & UI/UX designer & Formally diagnosed & -- & Letter reversal; line-skipping; visual blurring; mental exhaustion & Wall-of-text effect; small serif fonts \\
P20 & Female & 32 & PhD candidate & Researcher & Self-identified & -- & Phonological issues; slow decoding; anxiety; working memory deficit & Inconsistent formatting; verbose descriptions; time pressure \\
\bottomrule
\end{tabular}
} 
\end{table*}

\section{Full DFT-GEN Architecture}
To make this supplementary PDF self-contained, we include the architecture figure below.

\begin{figure*}[t]
\centering
\IfFileExists{framework.png}{%
\includegraphics[width=0.92\textwidth]{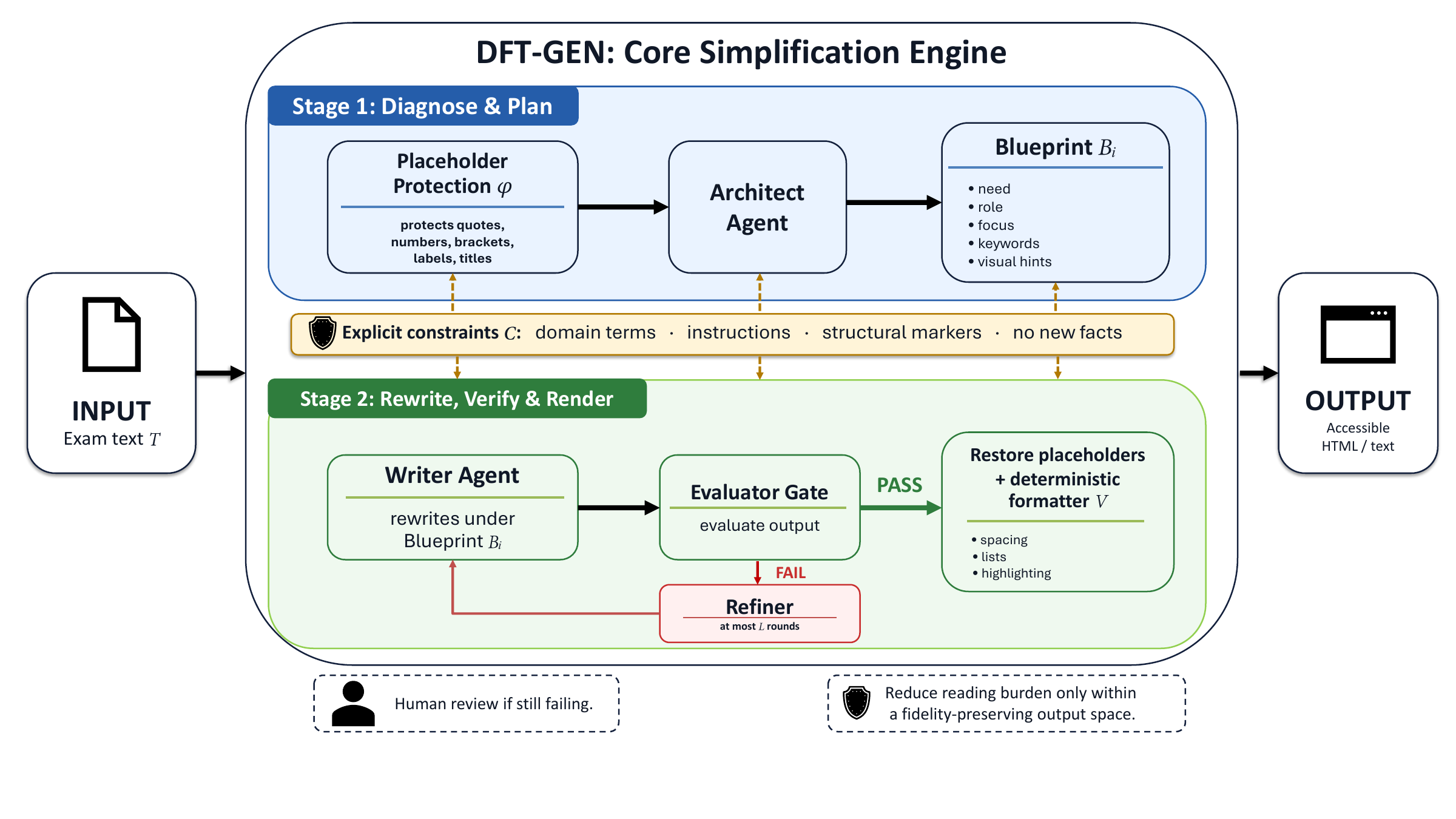}%
}{%
\fbox{\parbox{0.95\textwidth}{\vspace{0.5em}\centering
\textbf{DFT-GEN Architecture (missing \texttt{framework.png} in this build).}\\
Please include \texttt{framework.png} in the project root to render this figure.\vspace{0.5em}}}%
}
\caption{The ``Analyze-then-Reconstruct'' architecture of DFT-GEN. Stage~1 diagnoses each sentence to produce a transformation blueprint with preserved constraints. Stage~2 rewrites under the blueprint with placeholder protection and an Evaluator--Refiner loop, then restores placeholders and formats the final dyslexia-friendly output.}
\label{fig:architecture}
\end{figure*}

\section{Datasets}
\subsection{Overview and statistics}
We evaluate on two exam corpora: a Chinese Gaokao-style corpus and an English DSE-style corpus. Table~\ref{tab:dataset_stats_supp} summarizes dataset sizes and subject breakdown.

\begin{table}[t]
\centering
\scriptsize
\setlength{\tabcolsep}{3pt}
\caption{Dataset statistics for Chinese Gaokao and English DSE corpora.}
\label{tab:dataset_stats_supp}
\resizebox{\columnwidth}{!}{%
\begin{tabular}{lcc}
\toprule
\textbf{Statistic} & \textbf{Chinese (Gaokao)} & \textbf{English (DSE)} \\
\midrule
Total Items & 1,117 & 1,163 \\
History & 500 & 158 \\
Biology & 142 & 615 \\
Geography & 475 & 390 \\
Avg. Length (chars/words) & 1,129 chars & 86 words \\
\bottomrule
\end{tabular}
} 
\end{table}

\subsection{Construction details}
\paragraph{Chinese corpus.}
We constructed a high-complexity corpus derived from Gaokao and senior high school qualification exam questions. The resulting corpus contains 1,117 distinct question items spanning History (500), Geography (475), and Biology (142), with an average length of approximately 1,129 Chinese characters.

\paragraph{English corpus.}
We constructed an English-language corpus from DSE exam papers, containing 1,163 distinct question items spanning History (158), Biology (615), and Geography (390), with an average length of approximately 86 words.

\section{Models and baselines (details)}
We evaluated DFT-GEN with five state-of-the-art LLMs (DeepSeek-V3, Gemini-2.5-Pro, Qwen-3-Max, GPT-4o, Claude-4-Sonnet) and a commercial baseline (Dyslexia Helper). All models were accessed via APIs with temperature 0.3. Baselines include: layout-only (no rewriting), a neural ATS baseline (single-pass seq2seq rewriting), and a strong-prompt single-agent LLM baseline.

\section{Additional details moved out of the main paper}
\subsection{Prototype UI feature: multi-level information density (optional)}
We implemented an optional web UI feature that presents the \emph{same simplified content} at three information-density levels (Focus, Card, Full), differing only in the amount of text shown per card. This interface scaffold is a prototype-side rendering choice rather than a required component of the DFT-GEN generation pipeline. Its purpose is to reduce initial avoidance by allowing users to start with shorter segments and progressively reveal more context without changing task content.

\subsection{Reproducibility notes (high level)}
We provide an anonymized supplementary package that includes implementation details, experiment settings, and analysis artifacts sufficient to reproduce the reported numbers (including baselines). Due to third-party LLM usage policies and the risk of prompt leakage, we do not release full system prompts or few-shot exemplars; instead, we include prompt templates and a constraint checklist. Code and data will be released upon acceptance.

\section{Automatic evaluation and ablation tables}
The main paper includes the full automatic evaluation tables for both corpora and the five-stage ablation table. We omit duplicated copies here to keep the supplementary PDF concise.

\section{Sensitivity analysis}
\subsection{Impact of maximum refinement iterations \texorpdfstring{$L$}{L}}
\begin{table}[h]
\centering
\scriptsize
\setlength{\tabcolsep}{2.5pt}
\caption{Impact of maximum refinement iterations \texorpdfstring{$L$}{L} on performance.}
\label{tab:param_L_supp}
\begin{tabular}{lccccc}
\toprule
\textbf{$L$} & \textbf{DCFI} & \textbf{Eff.} & \textbf{Instr. Fid.} & \textbf{Punc. Fid.} & \textbf{Time (s)} \\
\midrule
1 & \textbf{0.728} & \textbf{0.760} & 1.000 & 0.600 & \textbf{22.4} \\
2 & 0.712 & 0.575 & 1.000 & \textbf{0.850} & 29.0 \\
3 & 0.652 & 0.515 & 1.000 & 0.700 & 35.2 \\
\bottomrule
\end{tabular}
\end{table}

\begin{figure}[h]
\centering
\IfFileExists{param_L_impact.png}{%
\includegraphics[width=0.9\linewidth]{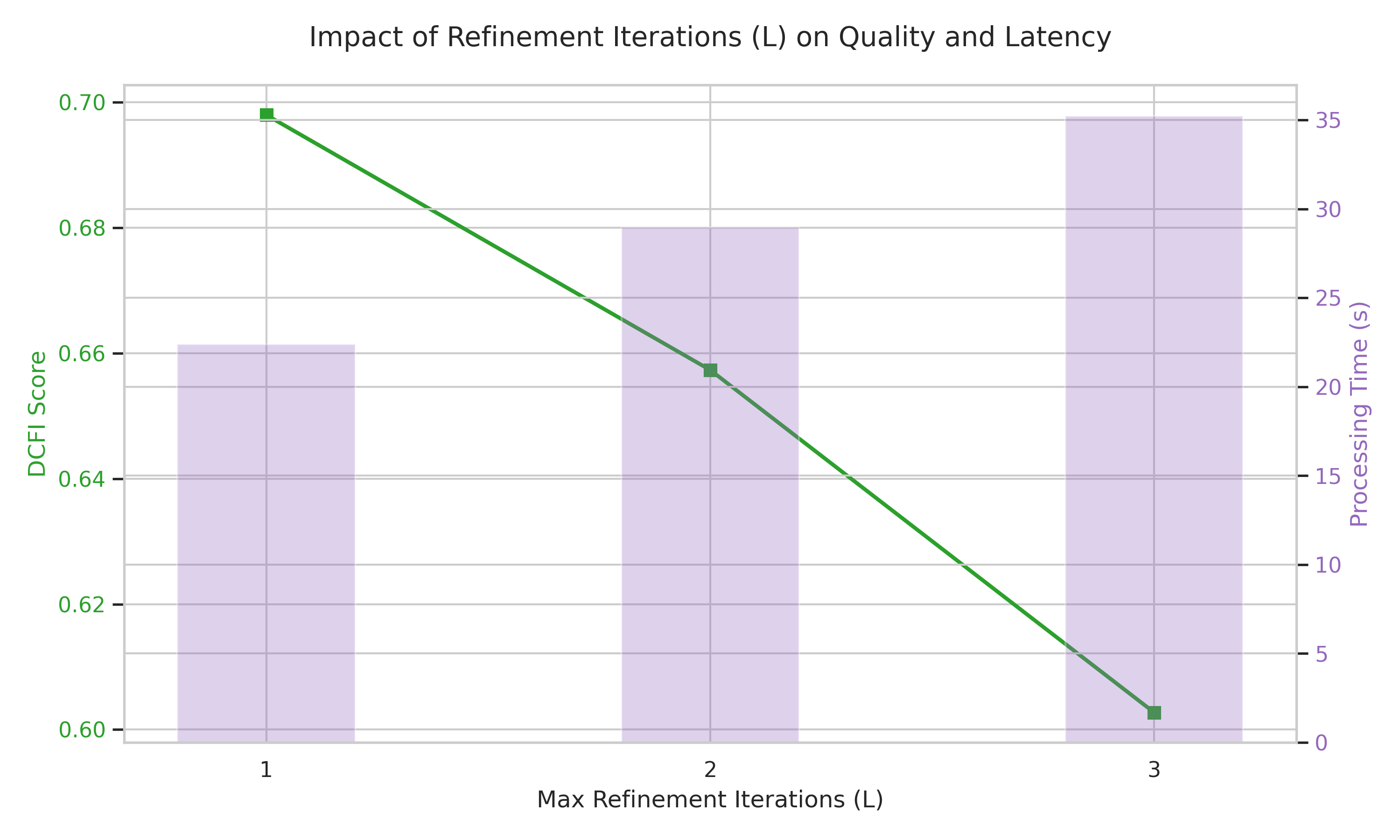}%
}{\fbox{\parbox{0.95\linewidth}{\centering Missing \texttt{param\_L\_impact.png}.}}}
\caption{Impact of refinement iterations ($L$). While DCFI peaks at $L=1$, structural fidelity (Punctual Fidelity) is maximized at $L=2$, motivating a conservative default for high-stakes use.}
\label{fig:param_L_supp}
\end{figure}

\subsection{Impact of DCFI component weights}
\begin{table}[h]
\centering
\caption{Impact of DCFI component weights. Weights are shown as $(w_e, w_c, w_i, w_p)$. Corr. denotes Spearman's $\rho$ with the default configuration.}
\label{tab:param_weights_supp}
\scriptsize
\begin{tabular}{lccc}
\toprule
\textbf{Configuration} & \textbf{Weights} & \textbf{DCFI} & \textbf{Corr.} \\
\midrule
Default & (0.30, 0.20, 0.25, 0.25) & 0.818 & 1.000 \\
Effectiveness-Heavy & (0.50, 0.15, 0.20, 0.15) & 0.764 & 0.357 \\
Fidelity-Heavy & (0.10, 0.30, 0.40, 0.20) & 0.821 & 0.929 \\
Punctual-Heavy & (0.10, 0.20, 0.30, 0.40) & 0.875 & 0.857 \\
Uniform & (0.25, 0.25, 0.25, 0.25) & 0.832 & 0.976 \\
\bottomrule
\end{tabular}
\end{table}

\begin{figure}[h]
\centering
\IfFileExists{param_weights_heatmap.png}{%
\includegraphics[width=0.9\linewidth]{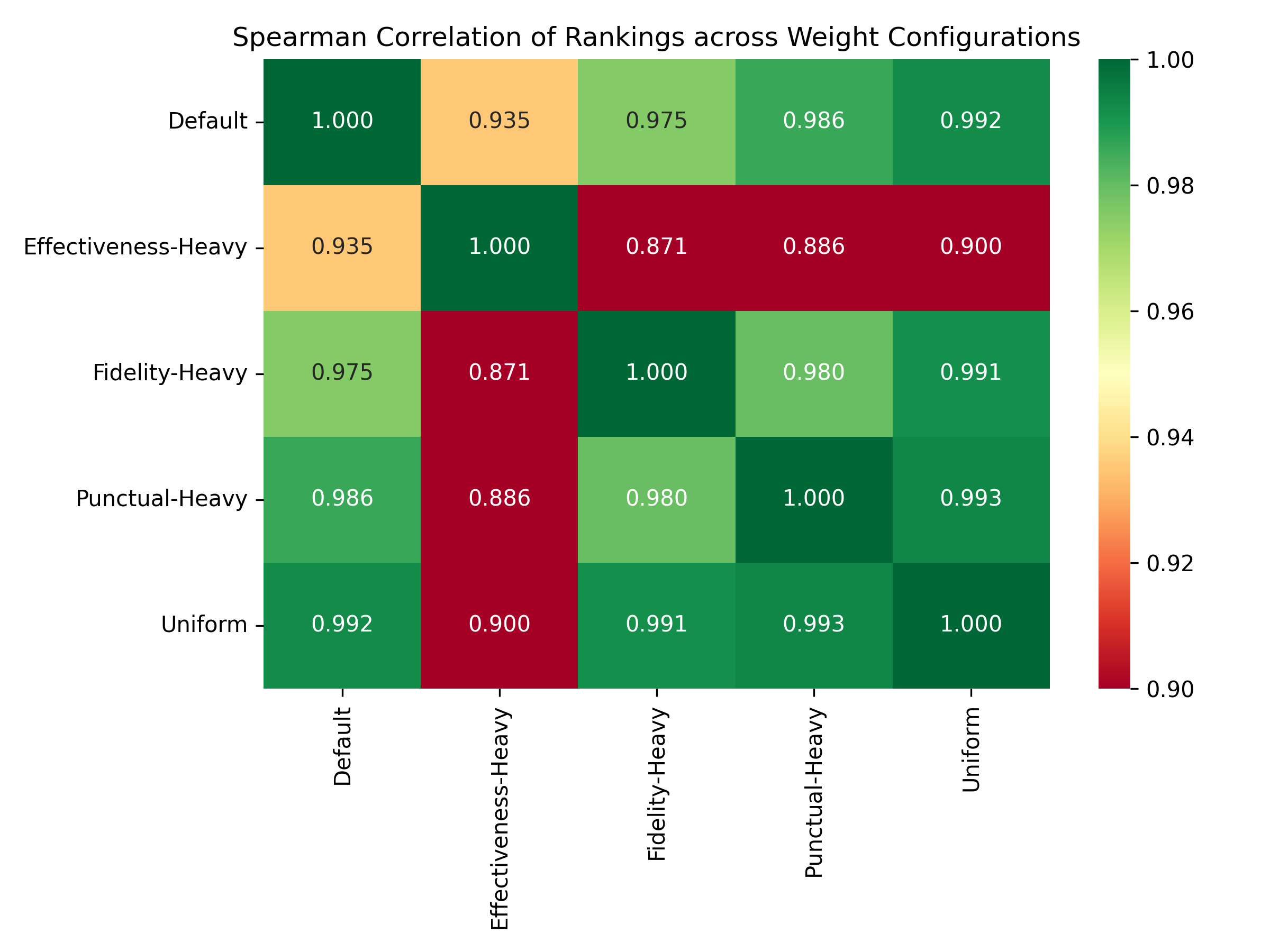}%
}{\fbox{\parbox{0.95\linewidth}{\centering Missing \texttt{param\_weights\_heatmap.png}.}}}
\caption{Spearman correlation matrix of model rankings across different DCFI weighting schemes.}
\label{fig:param_weights_supp}
\end{figure}

\section{Additional human-study details}
\subsection{Directional alignment: DCFI signals vs.\ human A/B outcomes}
\begin{table*}[t]
\centering
\scriptsize
\setlength{\tabcolsep}{2pt}
\caption{Directional alignment between DCFI signals and human A/B outcomes on two excerpts. Here $A$ is Original and $B$ is DFT-GEN. \textbf{Eff} denotes the effectiveness \emph{gain} signal $S_{\mathrm{eff}}$ (cognitive-load reduction relative to Original), hence $\textbf{Eff}_A=0$ by definition; human outcomes are group means for original vs.\ simplified.}
\label{tab:dcfi_human_alignment_supp}
\resizebox{\textwidth}{!}{%
\begin{tabular}{lcccccccccc}
\toprule
\textbf{Ex} & \textbf{DCFI$_A$} & \textbf{DCFI$_B$} & \textbf{Eff$_A$} & \textbf{Eff$_B$} & \textbf{Time$_A$} & \textbf{Time$_B$} & \textbf{Acc$_A$} & \textbf{Acc$_B$} & \textbf{Effort$_A$} & \textbf{Effort$_B$} \\
\midrule
Ex1 & 0.692 & 0.911 & 0.000 & 0.850 & 290.2 & 207.0 & 3.4 & 4.7 & 6.1 & 2.2 \\
Ex2 & 0.700 & 0.943 & 0.000 & 0.850 & 299.5 & 198.3 & 3.4 & 4.8 & 6.1 & 2.0 \\
\bottomrule
\end{tabular}
} 
\end{table*}

\subsection{Confidence ratings (two-condition run only)}
\begin{table}[h]
\centering
\footnotesize
\setlength{\tabcolsep}{4pt}
\caption{Confidence ratings (1--7) from an initial two-condition run only (n=10 per condition per excerpt).}
\label{tab:ab_confidence_supp}
\begin{tabular}{lccc}
\toprule
\textbf{Metric} & \textbf{Original} & \textbf{DFT-GEN} & \textbf{$p$} \\
\midrule
Confidence (1--7) (Ex1) & 3.3$\pm$1.2 & 5.9$\pm$1.0 & $2.4\times10^{-4}$ \\
Confidence (1--7) (Ex2) & 2.9$\pm$1.2 & 6.4$\pm$0.7 & $2.2\times10^{-5}$ \\
\bottomrule
\end{tabular}
\end{table}

\section{Reproducibility and data availability (anonymized)}
We provide an anonymized supplementary package that includes implementation details, experiment settings, and analysis artifacts sufficient to reproduce the reported numbers (including baselines). We also provide an anonymized code repository at \anoncodeurl that includes the simple text simplifier modules used in our ablations (\texttt{simple\_text\_simplifier.py}, \texttt{simple\_text\_simplifier\_en.py}) and the DCFI evaluation code (\texttt{evaluation.py}). Due to third-party LLM usage policies and the risk of prompt leakage, we do not release full system prompts or few-shot exemplars; instead, we include prompt templates (format/slots) and a constraint checklist.